\documentclass[11pt,letterpaper]{article}

\usepackage[english]{babel}
\usepackage{soul} % handling hyphenation
\usepackage[normalem]{ulem}  % underlining
\usepackage{setspace}  % control spacing between lines
\usepackage{xcolor}  % advanced colors, e.g., custom colors
\usepackage{color}  % color management
\usepackage{setspace}  % change line spacing
\usepackage[compact,md]{titlesec}  % customize section title formatting
\usepackage{textcomp}  % support for special text symbols, such as bullets
\usepackage[version=4]{mhchem}  % typesetting chemistry notation
\usepackage{siunitx}  % SI units

\usepackage[utf8]{inputenc}  % allows for unicode (UTF8) not just ASCII 
\usepackage{authblk}  % author and affiliation
\usepackage{blindtext}  % creates 'blind' content for testing
\usepackage{url}  % verbatim inclusion of URLs
\usepackage{etoolbox}  % advanced programming tools for classes/packages
\usepackage{tcolorbox}  % Adding colored box with text and equations
\usepackage[export]{adjustbox}  % adjust boxed content
\usepackage{subfiles}  % enable subfiles to be able to run standalone

\date{}

\makeatletter
\def\blfootnote{\gdef\@thefnmark{}\@footnotetext}
\makeatother

\usepackage{graphicx}  % including graphics
\usepackage{pdfpages}  % including external pdfs
\usepackage{pgfgantt}  % Gantt charts
\usepackage{caption}  % advanced captions
\usepackage{subcaption}  % captions for subfigures
\usepackage{rotating}  % rotate floats
\usepackage{wrapfig}  % enable wrapping text around figures/tables
\usepackage{tabularx}  % tables with adjustable-width columns
\usepackage{multirow}  % tables with cells spanning multiple rows
\usepackage{array}  % advanced arrays and tables
\usepackage{longtable}  % allow multi-page tables
\usepackage{makecell}  % advanced options for table layouts
\usepackage{colortbl}  % color for tables

\usepackage{amsmath}  % features for mathematical typesetting
\usepackage{amsfonts}  % extended fonts for mathematics
\usepackage{amssymb}  % extended symbols
\usepackage{amsthm}   % for theorems, proofs, lemmas, ...
\usepackage{esdiff}  % derivatives
\usepackage{siunitx}  % SI units
\usepackage{cancel}  % lines through math
\usepackage{bm}  % bold symbols in math

\usepackage{doi}  % create correct hyperlinks for DOIs
\usepackage[numbers]{natbib}  % to use doi (when using bibtex instead of biber engine)
\usepackage{hypernat} % makes hyperref anf natbib to work together
\usepackage{pdfsync}  % enable references between source and PDF
\usepackage{algorithm}  % floating algorithm environment
\usepackage{algpseudocode}  % actual algorithms
\usepackage{listings}  % typesetting code
\usepackage{pythonhighlight}

\usepackage{csquotes}

\titlespacing{\section}{0pt}{1.5ex}{1.5ex}
\titlespacing{\subsection}{0pt}{1.5ex}{1.5ex}
\titlespacing{\subsubsection}{0pt}{1.ex}{1.ex}

\titleformat{\section}[block]
{\center\normalfont\scshape}  % general formatting
{\textcolor{ucsd_blue}{\thesection}}  % the label and number
{0.5em}  % space between label/number and section name
{\color{ucsd_blue}}  % formatting applied just to section name
[]  % punctuation or other commands following section name

\titleformat{\subsection}[block]
{\normalfont}  % general formatting
{\textcolor{ucsd_blue}{\thesubsection}}  % the label and number
{0.5em}  % space between label/number and section name
{\color{ucsd_blue}}  % formatting applied just to section name
[]  % punctuation or other commands following section name

\titleformat{\subsubsection}[block]
{\normalfont}  % general formatting
{\textcolor{ucsd_blue}{\thesubsubsection}}  % the label and number
{0.5em}  % space between label/number and section name
{\color{ucsd_blue}}  % formatting applied just to section name
[]  % punctuation or other commands following section name

\definecolor{ucsd_blue}{RGB}{24, 43, 73}
\definecolor{ucsd_gold}{RGB}{198, 146, 20}
\definecolor{ucsd_blue_light}{RGB}{0, 98, 155}
\definecolor{ucsd_gold_light}{RGB}{255, 205, 0}
\definecolor{ucsd_gray}{RGB}{116, 118, 120}

\newcommand\Real{\mathbb{R}}  % set of real numbers
\newcommand\p\partial  % symbol in partial derivative
\newcommand\f\frac  % fraction

\begin{document}

\title{
    \textbf{
        modOpt: A modular development environment and library for optimization algorithms
    }
}

% \author{
%     Anugrah Jo Joshy\footnote{PhD Candidate, Department of Mechanical and Aerospace Engineering} \enspace and
%     John T. Hwang\footnote{Associate Professor, Department of Mechanical and Aerospace Engineering}
% }

\author[1,*]{
    Anugrah Jo Joshy
}
\author[1]{
    John T. Hwang 
}

\affil[1]{Department of Mechanical and Aerospace Engineering, 
          University of California San Diego, La Jolla, CA 92093, USA}
\affil[*]{Corresponding author: Anugrah Jo Joshy, ajoshy@ucsd.edu}

\renewcommand\Affilfont{\itshape\small}

% For bibliography using biber engine   
% \begin{refsection}
    \maketitle
    % \preprintheaderandtitle
    % {
    %     Anugrah Jo Joshy, and John T. Hwang. modOpt: A modular development environment and library for optimization algorithms. Structural and Multidisciplinary Optimization, 2024.
    % }
    % {10.25xx/1.J05xxxx}
    % {joshy2024modular}

    % \input{abstract}

    \begin{abstract}
    Applications of numerical optimization have appeared across a broad range of
    research fields, from finance and economics to the natural sciences and engineering.
    It is well known that the optimization techniques employed in each field
    are specialized to suit their problems.
    Recent advances in computing hardware and modeling software
    have given rise to new applications for numerical optimization.
    These new applications occasionally uncover bottlenecks in existing optimization 
    algorithms and necessitate further specialization of the algorithms.
    However, such specialization requires expert knowledge of the underlying 
    mathematical theory and the software implementation of existing algorithms.
    To address this challenge, we present modOpt, an open-source software
    framework that facilitates the construction of optimization algorithms from modules.
    The modular environment provided by modOpt enables developers to tailor 
    an existing algorithm for a new application by only altering the relevant modules.
    modOpt is designed as a platform to support students and beginner developers 
    in quickly learning and developing their own algorithms.
    With that aim, the entirety of the framework is written in Python, and it is
    well-documented, well-tested, and hosted open-source on GitHub.
    Several additional features are embedded into the framework  
    to assist both beginner and advanced developers.
    In addition to providing stock modules, the framework also includes fully 
    transparent implementations of pedagogical optimization algorithms in Python.
    To facilitate testing and benchmarking of new algorithms, the framework 
    features built-in visualization and recording capabilities, 
    interfaces to modeling frameworks such as OpenMDAO and CSDL, 
    interfaces to general-purpose optimization algorithms such as SNOPT and SLSQP,
    an interface to the CUTEst test problem set, etc.
    In this paper, we present the underlying software architecture of modOpt, 
    review its various features, 
    discuss several educational and performance-oriented algorithms within modOpt, 
    and present numerical studies illustrating its unique benefits.
    
    \end{abstract}
    
    % % \input{nomenclature}
    % \input{intro}
    \section{Introduction}
    \label{sec:intro}

    Optimization is the process of identifying the best choice among a
    set of alternatives.
    We are concerned here with nonlinear optimization,
    which is the solution of problems with nonlinear objective and constraint
    functions and continuous design variables.
    Such classes of problems frequently arise in several domains of research,
    from portfolio optimization in finance to aircraft design
    optimization in engineering.
    Numerical algorithms for solving these problems using a computer have
    been studied for more than half a century.

    It was evident from the beginning that no single optimization
    algorithm could efficiently solve all optimization problems,
    a notion later confirmed by the no free lunch (NFL) theorems \cite{wolpert1997no}.
    Naturally, algorithm development followed a divide-and-conquer
    approach, with researchers in each field devising dedicated algorithms 
    for specific classes of problems.
    Over time, many of these algorithms have matured,
    reaching their peak efficiency.
    The no free lunch (NFL) theorems \cite{wolpert1997no}, introduced by
    Wolpert in 1997, formally established that 
    `for any algorithm, any elevated performance over one class of problems 
    is offset by performance over another class'.
    A more practical interpretation of the NFL theorems is provided in \cite{ho2002simple},
    stating that 
    `a general-purpose universal optimization strategy is impossible, 
    and the only way one strategy can outperform another is if it is 
    specialized to the structure of the specific problem under consideration'.

    Computational capabilities have been gradually improving since the 1950s
    due to breakthroughs in hardware and software architectures.
    These developments catalyzed the creation of new categories
    of optimization problems and corresponding new classes of algorithms 
    for their solutions.
    However, recent advances in modeling software, the sudden rise
    of machine learning, and advances in high-performance computing (HPC)
    and graphics processing units (GPUs), among many others, have resulted 
    in an explosion in the number of optimization applications in the past decade.
    One notable example is the multitude of novel optimization applications
    made possible by NASA's OpenMDAO \cite{gray2019openmdao} modeling framework.
    The uniqueness of OpenMDAO lies in its ability to efficiently and 
    automatically compute total derivatives for large and complex computational models,
    provided that the partial derivatives for smaller components that constitute 
    the model are available.
    Through this capability, OpenMDAO has enabled a range of diverse applications, 
    including but not limited to aircraft design \cite{kao2015modular,hearn2016optimization},
    satellite design \cite{hwang2013large}, and 
    wind turbine design \cite{ning2014understanding}.

    New applications occasionally encounter bottlenecks within existing algorithms.
    For instance, quasi-Newton sequential quadratic programming (SQP) methods 
    for large-scale optimization problems struggle to converge when the
    Lagrangian Hessian cannot be represented effectively by a limited memory 
    method \cite{gill2015performance}.
    New developments in computational infrastructure open up 
    new avenues for improving existing algorithms.
    For example, the emergence of new software packages in popular programming languages like 
    Python and Julia, which enable efficient automatic differentiation (AD), 
    offers new possibilities.
    CasADI \cite{andersson2019casadi} and Jax \cite{jax2018github} are Python packages
    that can efficiently and automatically compute higher-order directional derivatives   
    using a mix of forward and adjoint sensitivity analysis.
    This capability can be leveraged for efficient optimization in new algorithms 
    utilizing Hessian- or Jacobian-vector products.
    Therefore, to promptly address emerging bottlenecks and capitalize on new
    innovations, optimization algorithms need to rapidly evolve.
    
    However, there are a few significant factors hindering this.
    The mathematical theory behind practical optimization algorithms is often quite complex, 
    making it tremendously time-consuming for new developers to fully understand before they can make specific adjustments. 
    Moreover, the software implementations of many widely used algorithms have undergone 
    decades of refinement and are primarily coded in low-level languages such as 
    C++ or Fortran.
    Documentation for contributing to some of the most popular 
    open-source algorithms, such as SLSQP, is limited or nonexistent.
    This lack of detailed guidelines makes it challenging for new contributors 
    to navigate the complexities of the codebase.
    Most internal optimization variables are often not readily accessible 
    from outside the optimizer, 
    which poses an additional challenge for those seeking to test and 
    fine-tune existing algorithms.
    In many instances, developers choose to write entirely new code from scratch 
    to circumvent these challenges.

    With the aim of lowering these barriers that hinder the development of 
    optimization algorithms, we introduce a software framework called modOpt, which stands 
    for `a modular development environment and library for optimization algorithms'.
    modOpt facilitates the construction of optimization algorithms from self-contained modules.
    Modules include line search algorithms, Hessian update algorithms, 
    and merit functions, among others.
    The modular environment empowers new developers to customize existing algorithms
    for new applications simply by altering the relevant modules.
    New ideas can be quickly tested by replacing as few as a single
    module within an existing algorithm,
    conserving valuable intellectual effort that would otherwise be spent on
    learning an existing algorithm and its implementation 
    in order to largely reproduce existing capabilities.
    modOpt also serves as a library of optimization algorithms with
    interfaces built to various gradient-based and gradient-free algorithms
    across different classes of optimization.
    This library, with a single, unified interface to multiple algorithms, 
    allows optimization practitioners to easily swap algorithms 
    when solving a problem for the first time.
    
    At a time when optimization topics are increasingly getting integrated into
    college curricula,
    the availability of beginner-friendly tools becomes crucial for helping
    students quickly familiarize themselves and experiment with 
    different optimization algorithms.
    modOpt's design philosophy, which involves breaking down large and complex 
    algorithms into more manageable modules, serves an equally important purpose
    of enabling students to incrementally build algorithms from scratch using modules or
    to specialize already available algorithms for their specific applications.
    Additionally, the library of algorithms is valuable for 
    projects where students need to apply optimization techniques. 
    Overall, modOpt will be a unique resource for instructors teaching 
    optimization courses.
    
    modOpt also includes an array of additional features to support users at all levels.
    In addition to providing stock modules, modOpt also includes
    fully transparent pedagogical optimization algorithms written in Python.
    To facilitate testing and benchmarking of new algorithms, the framework 
    features built-in visualization and recording capabilities, 
    interfaces to modeling frameworks such as OpenMDAO and CSDL 
    \cite{gandarillas2024graph},
    and an interface to the CUTEst \cite{gould2015cutest} problem set.
    modOpt also has built-in modeling capabilities for problems ranging from
    those without derivatives to those with full or directional second-order derivatives.
    The framework can accommodate algorithms for a wide range of problem classes,
    including but not limited to quadratic programming (QP), convex programming,
    and nonlinear programming (NLP).
    modOpt is entirely written in Python, and is well-documented, well-tested, and
    published open-source on GitHub.

    The remainder of this paper is organized as follows.
    In Section \ref{sec:bg}, we discuss the formulation of optimization problems, 
    the general workflow, and provide a brief survey of popular tools employed in optimization.
    In Section \ref{sec:arch}, we present the software design of modOpt,
    including simple scripts illustrating its usage.
    Section \ref{sec:opt} serves as a quick reference guide for 
    the optimization algorithms implemented in or interfaced with modOpt.
    In Section \ref{sec:egs}, we present various numerical studies uniquely enabled by modOpt.
    Section \ref{sec:conclusion} concludes the paper with closing remarks.

    \section{Background}
\label{sec:bg}

In this section, we present the general optimization workflow and a brief overview
of commonly used tools in optimization. 
\label{sec:background}
\subsection{Optimization formulation and workflow}
The focus in this paper will be on nonlinear programming (NLP) problems which, in general, 
can be formulated as:
\begin{equation}
    \begin{array}{r l}
        \text{minimize}        & f(x)                 \\
        \text{with respect to} & x \\
        \text{subject to}      & c(x)\geq0, \\
    \end{array}
    \label{eq:opt}
\end{equation}
where $x\in\Real^n$ represents the vector of optimization variables, 
$f:\Real^n\to\Real$ represents the scalar objective function, and
$c:\Real^n\to\Real^m$ represents the vector-valued constraint function.
Maximization problems and problems with equality constraints can be 
transformed into the general form above.
Nonlinear programs are optimization problems with nonlinear and smooth objective 
and constraint functions defined over continuous optimization variables.

\begin{figure}[ht]
    \centering
    \includegraphics[width=0.3\linewidth]{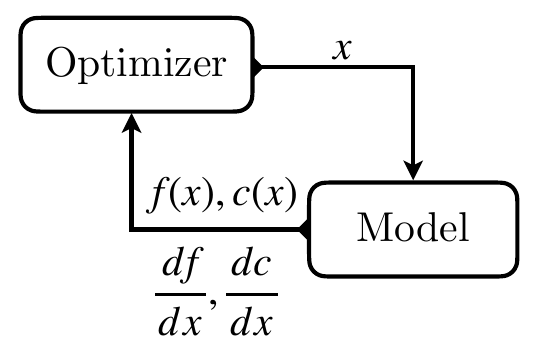}
    \caption{\textbf{Optimization workflow}
        % \normalfont{}
    }
    \label{fig:workflow}
\end{figure}

In practice, numerical optimization problems are solved by programming a 
computational model and linking it to a general-purpose optimization algorithm,
also called an optimizer (shown in Fig. \ref{fig:workflow}).
Given the optimization variable vector $x$ as input, the model computes the 
optimization functions $f(x)$ and $c(x)$, 
and, in some cases, their derivatives $df/dx$ and $dc/dx$.
The optimizer iteratively evaluates the model at different $x$ values until it finds a
sufficiently optimal solution $x^*$ to the problem.

When using a gradient-free optimizer, derivatives are not required,
and the model computes only the function values $f(x)$ and $c(x)$.
However, gradient-free algorithms scale poorly with the number of optimization variables $n$,
making gradient-based algorithms more suitable for problems with large $n$.
Some gradient-based algorithms can further benefit from access to
second derivatives, in which case the models also compute the second derivatives.
Unlike many gradient-free algorithms 
that are capable of finding global minima, 
most gradient-based algorithms are limited to 
identifying local minima in optimization problems.

Algebraic modeling languages (AMLs) are designed to formulate
optimization models using standard mathematical operations and expressions.
They allow users to define the objective and constraints for a problem
using algebraic expressions.
AMLs often come equipped with interfaces to optimization algorithms 
for solving the modeled problems.
AMPL \cite{fourer2003amplmain}, GAMS \cite{rosenthal2004gamsmain}, Pyomo \cite{hart2011pyomo},
JuMP \cite{dunning2017jump} and CasADi \cite{andersson2019casadi} are some of the most 
popular modeling languages for optimization.
Jax \cite{jax2018github} is a machine-learning-focused, high-performance 
Python library that has recently seen significant adoption as a modeling 
language for optimization.

The aforementioned AMLs do not explicitly support the modeling of 
large and complex multidisciplinary systems, 
such as those encountered in aircraft or satellite design.
OpenMDAO \cite{gray2019openmdao} and CSDL \cite{gandarillas2024graph} are
Python libraries that address this limitation by enabling the construction of
large and complex models from modular components.
OpenMDAO efficiently automates total derivative computation using partial
derivatives for the components of a decomposed model.
Numerous real-world design optimization studies have been successfully carried out
using OpenMDAO
\cite{kao2015modular, hearn2016optimization, hwang2013large, ning2014understanding}.
CSDL, on the other hand, automatically computes the total derivatives of 
the objective and constraints in large models using 
automatic differentiation (AD) and implicit differentiation techniques.
This capability has been instrumental in enabling some of the most comprehensive 
aircraft design optimizations to date \cite{sarojini2023large, ruh2024large}.

\subsection{General-purpose optimization algorithms}
General-purpose optimization algorithms, also known as optimizers, are software
designed to address specific classes of optimization problems.
These numerical algorithms are often implemented in low-level programming 
languages such as C++ or Fortran to maximize computational efficiency.
Examples of widely used gradient-based optimizers for nonlinear programming (NLP) include
SLSQP \cite{kraft1988software},  SNOPT \cite{gill2005snopt}, 
IPOPT \cite{wachter2006implementation}, LOQO \cite{vanderbei1999loqo}, and 
KNITRO \cite{byrd1999knitro}.
SNOPT and SLSQP are sequential quadratic programming (SQP) algorithms 
whereas IPOPT, LOQO, and KNITRO are interior point (IP) algorithms.
All the aforementioned algorithms can handle problems with hundreds of optimization
variables and constraints.
SLSQP and IPOPT are open-source and have facilitated significant research, 
making them among the most popular choices for nonlinear programming.
Some general-purpose optimizers are more specialized for certain applications than others. 
For instance, ParOpt \cite{chin2019paropt} is an interior point algorithm 
specifically tailored for topology optimization.

There is a multitude of algorithms targeting other classes of optimization problems.
Linear programming, quadratic programming, convex programming, 
and stochastic programming, to name a few, represent some of these diverse classes.
In the area of gradient-free optimization, 
particle-swarm optimization \cite{kennedy1995particle}, 
genetic algorithms \cite{goldberg2013genetic}, 
and simulated annealing \cite{kirkpatrick1983simulated} are 
some of the popular methods.

\subsection{Optimizer libraries}
Optimization algorithms are implemented in different programming languages, 
each requiring a distinct formulation to specify the optimization problem.
For instance, the formulation of constraints can vary significantly 
among different optimizers,
affecting how users specify constraint bounds, linear or nonlinear constraints,
and equality or inequality constraints.
Such differences impose an additional burden on practitioners who must 
develop optimizer-specific interfaces when seeking the best optimizer 
for their applications.
Optimization algorithm developers also encounter similar challenges when 
benchmarking newly developed algorithms against existing ones.
To address these issues, optimizer libraries have been developed.

Optimizer libraries provide a unified interface for multiple optimization algorithms,
allowing users to seamlessly switch between different optimizers for a given problem.
These libraries typically include source code for open-source algorithms and 
provide interfaces for proprietary algorithms.
Additionally, optimizer libraries may offer utilities for
modeling, scaling, recording, warm-starting, visualization, 
benchmarking, and post-processing.
Scipy.minimize \cite{virtanen2020scipy}, 
pyOpt \cite{perez2012pyopt}, and
pyOptSparse \cite{wu2020pyoptsparse}
are among the most popular optimizer libraries in Python for nonlinear programming.
Examples of optimizer libraries for other classes of problems include
CVXOPT \cite{andersen2020cvxopt} for convex programming and 
qpsolvers \cite{qpsolvers2024} for quadratic programming.

Although optimizer libraries provide a common interface to different optimizers,
to the best of our knowledge, none of them provide explicit support for 
optimizer development.
pyOpt and pyOptSparse, like many other libraries, allow the integration of
new optimizers into their frameworks but lack dedicated tools for developing 
new optimization algorithms.
Moreover, these frameworks are restricted to certain algorithms by design.
For example, they do not support directional derivatives or second derivatives
in models or optimizers.
SciPy's \textit{minimize} module supports a variety of optimization algorithms. 
However, it requires developers to have advanced programming knowledge and 
expects them to independently navigate its vast and complex codebase.

The modOpt software framework introduced in this paper 
functions as an optimizer library, offering a suite of built-in instructional algorithms 
and interfaces to several high-performance optimization algorithms.
modOpt incorporates all the optimization utilities previously mentioned and 
includes a diverse range of algorithms, such as those for quadratic programming, 
nonlinear programming, and convex programming, among others.
modOpt equips its users and developers with dedicated 
Python tools for customizing existing optimizers and building new ones.
It also provides interfaces to multiple modeling languages, such as 
Jax, CasADi, CSDL, and OpenMDAO, along with a built-in, intuitive class 
for modeling simpler problems.

    \section{Software architecture}
\label{sec:arch}

As discussed in Sec. \ref{sec:bg}, forming and solving an optimization problem 
requires both a model and an optimizer (see Fig. \ref{fig:workflow}) . 
In modOpt, we adopt an object-oriented approach to facilitate the 
specification and coupling of models and optimizers. 
Models are defined using the \texttt{Problem} or \texttt{ProblemLite} classes, 
while optimizers are defined using the \texttt{Optimizer} class. 
The \texttt{Problem} and \texttt{ProblemLite} classes provide attributes and methods 
that establish a unified interface to all optimizers, 
enabling them to evaluate models consistently.
This design ensures independence between models and optimizers, 
allowing users to interchange models for a given optimizer and vice-versa. 
The major classes that facilitate this approach, along with their relationships, 
are depicted in the UML class diagram in Fig. \ref{fig:uml}.
In this section, we will briefly review these classes, present simple optimization scripts,
and summarize the features enabled by modOpt.

\subsection{\texttt{Problem} class}
\label{sec:Problem}
The \texttt{Problem} class encapsulates the optimization problem definition
and is designed as an abstract class.
Users formulate their specific problem by subclassing \texttt{Problem}.
The attributes and methods accessed by the optimizer are automatically generated 
based on how the user defines the subclass.

The subclass definition begins with the \texttt{initialize} method, which 
assigns a name to the problem and declares any model-specific options.
The \texttt{setup} method follows, adding the objective, design variables, 
and constraints using the \texttt{add\_objective}, \texttt{add\_design\_variables}, 
and \texttt{add\_constraints} methods, respectively.
Note that within modOpt, `design variables' are synonymous with `optimization variables'.
Users can also specify scaling for all variables and functions, as well as 
lower and upper bounds for variables and constraints, while adding them to the problem.

\newpage 

\begin{figure}[ht]
    \centering
    \includegraphics[width=0.85\linewidth]{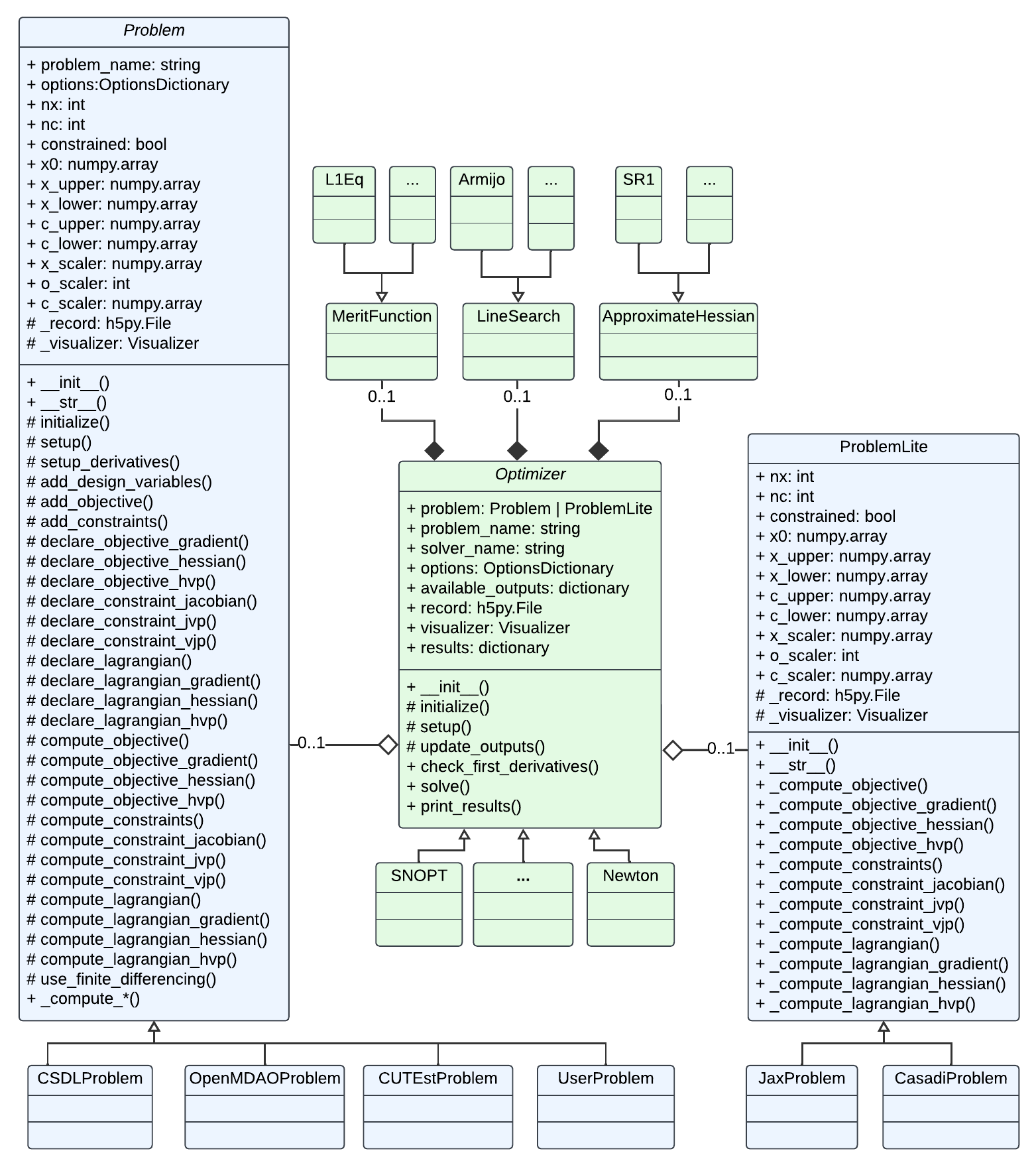}
    \caption{\textbf{UML class diagram for modOpt}
        % \normalfont{}
    }
    \label{fig:uml}
\end{figure}

For gradient-based optimizers, users must define the \texttt{setup\_derivatives} method 
to declare the derivatives they will provide later.
Depending on the problem and optimizer, these derivatives may include 
objective gradients, constraint Jacobians, or objective/Lagrangian Hessians.
Optimizers that can directly use the Lagrangian and its derivatives benefit 
from users declaring these as well.
Similarly, users can also declare matrix-vector products (Jacobian-vector products (JVPs), 
vector-Jacobian products (VJPs), objective/Lagrangian Hessian-vector products (HVPs)) 
when using optimizers that can leverage them.

For a problem with distinct design variables and constraints, 
these quantities can be declared separately. 
They also require separate derivative declarations.
The \texttt{Problem} class employs efficient array management techniques to
handle sparsity in constraint Jacobians and Hessians, 
allowing for computational and memory savings when sub-Jacobians or 
sub-Hessians with different sparsity patterns are declared individually.

Once all functions are declared, users must define the methods to compute 
these functions, even if they are constants.
This requirement ensures consistency and prevents users from inadvertently 
omitting necessary definitions.
For example, if an objective and its gradient are declared, the user must define 
the \texttt{compute\_objective} and \texttt{compute\_objective\_gradient} methods.
Method names are kept 
verbose instead of using abbreviations, 
to be maximally explicit and to avoid ambiguities.
The full list of optimization functions supported by the \texttt{Problem} class is 
listed in Fig. \ref{fig:uml}.
If certain derivatives are challenging to formulate, users can invoke the 
\texttt{use\_finite\_differencing} method within the corresponding `compute' method 
to automatically apply a first-order finite difference approximation.

When instantiated, a subclass object generates the necessary attributes for the optimizer, 
such as the initial guess vector, scaling vectors, and bound vectors
corresponding to the concatenated vectors for the design variables and constraints.
The \texttt{Problem} class also includes methods that wrap the user-defined 
`compute' methods for the optimizer, applying the necessary scaling beforehand.
These wrapped `compute' methods are prefixed with an underscore. 
When recording and visualizing, these wrapper methods additionally update 
the \texttt{record} and the \texttt{Visualizer} objects with the latest values 
from each `compute' call.

Strong interfaces to various modeling languages and test-suite problems are 
established through subclasses of the \texttt{Problem} class.
For instance, \texttt{OpenMDAOProblem}, \texttt{CSDLProblem}, 
and \texttt{CUTEstProblem} inherit from \texttt{Problem}.
These subclasses serve as wrappers that transform the models written in their 
respective languages into models suitable for the optimizers in modOpt.

The following code snippet demonstrates how to define the problem
\begin{align}
\underset{x \in \mathbb{R}^2}{\text{minimize}} & \quad x_1^2 + x_2^2 \notag \\
\text{subject to} & \quad x_1\geq0, \label{eq:example} \\
                  & \quad x_1+x_2=1,  \notag\\
                  & \quad x_1-x_2\geq1, \notag
\end{align}
using the \texttt{Problem} class.

\begin{python}
import numpy as np
from modopt import Problem

class Quadratic(Problem):
    def initialize(self, ):
        self.problem_name = 'quadratic'

    def setup(self):
        self.add_design_variables('x',
                                  shape=(2, ),
                                  lower=np.array([0., -np.inf]),
                                  upper=np.array([np.inf, np.inf]),
                                  vals=np.array([500., 5.]))
        self.add_objective('f')
        self.add_constraints('c',
                             shape=(2, ),
                             lower=np.array([1., 1.]),
                             upper=np.array([1., np.inf]),
                             equals=None,)

    def setup_derivatives(self):
        self.declare_objective_gradient(wrt='x', vals=None)
        self.declare_constraint_jacobian(of='c', wrt='x',
                                         vals=np.array([[1.,1.],[1.,-1.]]))

    def compute_objective(self, dvs, obj):
        x = dvs['x']
        obj['f'] = np.sum(x**2)

    def compute_objective_gradient(self, dvs, grad):
        grad['x'] = 2 * dvs['x']

    def compute_constraints(self, dvs, cons):
        x   = dvs['x']
        con = cons['c']
        con[0] = x[0] + x[1]
        con[1] = x[0] - x[1]

    def compute_constraint_jacobian(self, dvs, jac):
        pass
\end{python}

\subsection{\texttt{ProblemLite} class}
\label{sec:ProblemLite}

Many optimization problems, e.g., academic test problems
designed to evaluate specific aspects of optimization algorithms,
are simple and do not require the advanced array 
management techniques provided by the \texttt{Problem} class
discussed previously.
For such problems, it is often more efficient for users to define 
the optimization functions in pure Python and provide these functions, 
along with constants such as the initial guess, bounds, and scalers,
directly to modOpt.
Furthermore, some practitioners may prefer to use the optimizers in modOpt 
without involving the \texttt{Problem} class, 
particularly if they have already modeled their optimization functions outside of modOpt.

In these scenarios, subclassing the \texttt{Problem} class
can be excessive and result in unnecessarily verbose code. 
The \texttt{ProblemLite} class is designed to address this situation.
Unlike \texttt{Problem}, \texttt{ProblemLite} is a concrete class that 
can be instantiated directly, without requiring any subclassing.
\texttt{ProblemLite} objects act as containers for 
optimization constants and functions defined externally,
streamlining the process for users primarily focused on using the optimizers.
Additionally, \texttt{ProblemLite} abstracts away the more advanced 
object-oriented programming concepts, making it more accessible to beginners.

Despite its simplicity, \texttt{ProblemLite} replicates the same interface 
as \texttt{Problem} when interacting with the optimizer (see Fig. \ref{fig:uml}).
It supports recording and visualization, and 
can automatically generate derivatives using finite differences 
if the derivatives requested by the optimizer are not provided.
It also facilitates lightweight interfaces to
optimization functions implemented in other modeling languages.
For example, the \texttt{JaxProblem} and \texttt{CasadiProblem} in modOpt are 
implemented as subclasses of \texttt{ProblemLite}.

The code below demonstrates the use of the \texttt{ProblemLite} class
for the same problem presented in Eq. (\ref{eq:example}), 
illustrating the reduction in code complexity and 
the near elimination of object-oriented programming.

\begin{python}
import numpy as np
from modopt import ProblemLite

name = 'quadratic'
x0 = np.array([500., 5.])
xl = np.array([0., -np.inf])
cl = np.array([1., 1.]),
cu = np.array([1., np.inf]),

obj  = lambda x: np.sum(x**2)
grad = lambda x: 2 * x
con  = lambda x: np.array([x[0]+x[1], x[0]-x[1]])
jac  = lambda x: np.array([[1., 1.], [1., -1.]])

prob = ProblemLite(name=name, x0=x0, xl=xl, cl=cl, cu=cu,
                   obj=obj, grad=grad, con=con, jac=jac)
\end{python}

\subsection{\texttt{Optimizer} class}
The \texttt{Optimizer} class 
performs the role of encapsulating optimization algorithms within modOpt. 
Its design shares many similarities with the \texttt{Problem} class.
Like \texttt{Problem}, the \texttt{Optimizer} class is abstract, 
requiring users to create specific optimizers through subclassing.
All optimizers in modOpt must inherit from the \texttt{Optimizer} class.
Existing algorithms, such as SNOPT and IPOPT, are integrated into modOpt 
as subclasses of the \texttt{Optimizer} class.

Every optimizer in modOpt must be instantiated with a problem object, 
which should be derived from the \texttt{Problem} or \texttt{ProblemLite} classes, 
as detailed in the previous subsections.
The \texttt{Optimizer} base class provides essential tools 
for recording, visualization, and hot-starting an optimization.
The \texttt{record} attribute manages the systematic recording of the optimization, 
while the \texttt{visualizer} attribute enables real-time visualization of 
the optimization process.
The \texttt{Optimizer} base class also implements a \texttt{check\_first\_derivatives} 
method to verify the correctness of the user-defined derivatives 
in the provided problem object.

Subclasses of \texttt{Optimizer} must implement an \texttt{initialize} method
that sets the \texttt{solver\_name} and declares any optimizer-specific options.
Developers are required to define the \texttt{available\_outputs} attribute within 
the \texttt{initialize} method.
This attribute specifies the data that the optimizer will provide 
after each iteration of the algorithm by calling the \texttt{update\_outputs}
method.
Developers can also optionally define a \texttt{setup} method to handle any pre-processing
of the problem data and configuration of the optimizer's modules.

The core of an optimizer in modOpt lies in the \texttt{solve} method.
This method implements the numerical algorithm and 
iteratively calls the `\_compute' methods from the problem object.
Upon completion of the optimization, 
the \texttt{solve} method should assign a \texttt{results} attribute 
that holds the optimization results in the form of a dictionary.
Developers may optionally implement a \texttt{print\_results} method 
to override the default implementation provided by the base class and
customize the presentation of the results.

New optimizers can leverage the modules already available within modOpt.
These modules include Hessian approximation algorithms, 
merit functions, and line searches.
For algorithms that require QP solutions, developers can also utilize
several QP solvers that are interfaced with modOpt.
Merit functions available in modOpt include $l_1$, $l_2$ and $l_{\infty}$ penalty functions,
as well as Lagrangian, modified Lagrangian, and augmented Lagrangian functions.
modOpt includes backtracking line searches enforcing Armijo and Wolfe conditions, 
along with interfaces to line searches from SciPy and MINPACK \cite{more1980user}.
Hessian approximations implemented in modOpt include 
the BFGS (Broyden--Fletcher--Goldfarb--Shanno), DFP (Davidon--Fletcher--Powell), 
PSB (Powell Symmetric Broyden), symmetric rank-one, 
Broyden , Broyden first, and Broyden class algorithms.

The following example shows the implementation of the BFGS algorithm for
unconstrained optimization, employing modules for
approximating Hessians and performing line searches.

\begin{python}
import numpy as np
import time
from modopt import Optimizer
from modopt.line_search_algorithms import ScipyLS
from modopt.approximate_hessians import BFGS

class QuasiNewton(Optimizer):
    def initialize(self):
        self.solver_name = 'bfgs'
        self.obj  = self.problem._compute_objective
        self.grad = self.problem._compute_objective_gradient
        self.options.declare('maxiter', default=500, types=int)
        self.options.declare('opt_tol', types=float, default=1e-6)
        self.options.declare('readable_outputs', types=list, default=[])
        self.available_outputs = {
            'itr': int,
            'obj': float,
            # Declare shape for an array output
            'x': (float, (self.problem.nx, )),
            'opt': float,
        }

    def setup(self):
        self.LS = ScipyLS(f=self.obj, g=self.grad)
        self.QN = BFGS(nx=self.problem.nx)

    def solve(self):
        opt_tol = self.options['opt_tol']
        maxiter = self.options['maxiter']

        start_time = time.time()
        x_k = self.problem.x0 * 1.
        f_k = self.obj(x_k)
        g_k = self.grad(x_k)
        opt = np.linalg.norm(g_k)
        itr  = 0

        # Updating outputs from the 0-th iteration
        self.update_outputs(itr=itr,
                            x=x_k,
                            obj=f_k,
                            opt=opt)

        while (opt > opt_tol and itr < maxiter):
            itr += 1
            # Hessian approximation
            B_k = self.QN.B_k
            # Compute the search direction toward the next iterate
            p_k = np.linalg.solve(B_k, -g_k)
            # Compute the step length along the search direction via a line search
            alpha, f_k, g_new, slope_new, new_f_evals, new_g_evals, converged = 
                self.LS.search(x=x_k, p=p_k, f0=f_k, g0=g_k)
                
            d_k   = alpha * p_k
            x_k  += d_k
            g_up  = self.grad(x_k)
            w_k   = g_up - g_k
            g_k   = g_up
            opt   = np.linalg.norm(g_k)          
            self.QN.update(d_k, w_k)

            # Updating outputs at the end of each iteration
            self.update_outputs(itr=itr,
                                x=x_k,
                                obj=f_k,
                                opt=opt)

        self.results = {
            'x': x_k, 
            'objective': f_k, 
            'optimality': opt, 
            'niter': itr, 
            'time': time.time() - start_time,
            'converged': opt <= opt_tol,
            }
        # Run post-processing from the Optimizer() base class
        self.run_post_processing() 
        return self.results

\end{python}

\subsection{Software features}
This section presents a summary of the features offered by modOpt
to support modeling, optimization, and algorithm development.

\paragraph{\textbf{Unified interface between models and optimizers.}}

The object-oriented architecture of modOpt was designed with the 
intent to accommodate a diverse range of models and optimizers,
resulting in a singular, streamlined interface between the two.
This architecture empowers users to easily swap between different 
models and optimizers, a feature that proves particularly useful 
in cases where an application is being optimized for the first time
which requires trying multiple optimizers, or when developers need 
to test their new algorithms across various model test cases.
Moreover, newly integrated optimizers can instantly leverage 
additional functionalities such as scaling, recording, hot-starting,
and visualization, all readily available within
modOpt's base classes without any additional configuration.

To showcase the unified interface, 
the script below optimizes the problem defined previously in 
Section \ref{sec:Problem}, using the \texttt{PySLSQP} optimizer.
In this example, we activate the recording feature and visualize 
both the objective and the variable $x_1$,
throughout the optimization.
\begin{python}
from modopt import PySLSQP

optimizer = PySLSQP(Quadratic(), 
                    solver_options={'acc':1e-6}, 
                    recording=True,
                    visualize=['obj', 'x[0]'])
results = optimizer.solve()

\end{python}
Furthermore, to reduce dependence on object-oriented paradigms, 
modOpt offers a procedural interface, enabling users 
to bypass object-oriented structures entirely. 
The \texttt{optimize} function allows direct
optimization without requiring object instantiation. 
The following code demonstrates its usage by optimizing the
\texttt{ProblemLite} model from Section \ref{sec:ProblemLite}.

\begin{python}
from modopt import optimize

results = optimize(prob,
                   solver='PySLSQP',
                   solver_options={'acc':1e-6}, 
                   recording=True,
                   visualize=['obj', 'x[0]'])

\end{python}

\paragraph{\textbf{Flexibility in modeling.}}
In addition to the modeling facilitated by the \texttt{Problem} and 
\texttt{ProblemLite} classes, modOpt supports models developed
in CSDL, OpenMDAO, Jax, and CasADi.
This flexibility is particularly advantageous for students and 
practitioners when developing new models for optimization.
Jax and CasADi are well-suited for models represented by extensive
explicit analytic equations, as well as linear systems.
While CasADi’s API is more complex than Jax's, 
CasADi offers comprehensive support for sparse matrices 
and provides solvers that are optimized for domain-specific 
problems like optimal control.
Despite both Jax and CasADi employing automatic differentiation (AD) 
and implicit differentiation, Jax tends to be more efficient in memory
and time when handling large second-order models with implicit solvers.
Moreover, Jax supports GPU acceleration.

OpenMDAO and CSDL, on the other hand, are specifically designed 
for modeling large, multidisciplinary systems with complex interactions
between subsystems.
Due to the added framework overhead needed to support complex modeling 
requirements, they are generally less efficient than Jax or CasADi.
Modeling in OpenMDAO requires more planning and effort to finely break down 
the system-level model into smaller components, as it requires 
users to supply the partial derivatives for these components.
CSDL, by contrast, fully automates derivative computation using AD
and, unlike OpenMDAO, is capable of generating higher-order derivatives.
Additionally, CSDL’s backend leverages Jax, providing higher
computational efficiency than OpenMDAO, with GPU acceleration.
However, CSDL can incur additional overheads from AD, which can make it less 
memory-efficient than OpenMDAO for complex, large-scale models involving 
\textit{for loops}.

\paragraph{\textbf{Versatile library.}}
The library of optimizers in modOpt is not confined to methods for any
particular class of problems, unlike prior libraries.
This flexibility is essential to accommodate a diverse array of 
optimizers, which is invaluable for both research and education.
The library currently includes a broad range of both performant 
and educational algorithms.
Additionally, it inherently supports algorithms that can utilize first 
and second derivatives, as well as directional derivatives represented 
as matrix-vector products.
The library is also compatible with models and algorithms 
incorporating sparse or dense Hessians and constraint Jacobians.

\paragraph{\textbf{Modular development environment.}}
The modOpt framework introduces a novel object-oriented paradigm that 
allows developers to construct new optimization algorithms in a modular 
and extensible manner.
Implemented in Python, this framework enhances transparency of optimizers 
by providing users with greater access to optimization data, 
simplifying the process of tuning optimizers and refining models.
The reference algorithms and stock modules within the library streamline
the development cycle, significantly reducing the time required to prototype 
and test new algorithms. 
Moreover, its modular and transparent structure aids students and 
less experienced developers in implementing complex algorithms 
through reusable components, while also promoting efficient learning 
and practical application of optimization techniques.

\paragraph{\textbf{Numerical differentiation.}}
When exact derivatives are difficult to derive or implement, 
the \texttt{Problem} and \texttt{ProblemLite} classes provide a means 
to approximate first and second derivatives using first-order 
finite differences. 
Additionally, these classes efficiently compute numerical derivatives
for matrix-vector products such as JVPs, VJPs, and HVPs. 
While these approximations are available, it is preferable to 
utilize analytical derivatives to ensure computational accuracy and 
efficiency. 
First-order finite differencing requires $n$ extra function evaluations
per derivative and is prone to subtractive cancellation errors.

\paragraph{\textbf{Scaling.}}
In practical applications, it is often necessary to scale 
optimization variables and functions for achieving convergence.
However, this requires users to manually adjust the initial
guesses, variable and constraint bounds, and function and derivative
definitions.
modOpt alleviates this need by decoupling scaling from the 
original problem definition, allowing users to specify the scaling 
parameters separately when rerunning an optimization with a new scaling.
The framework automatically applies these scaling factors 
to the initial guesses, bounds, functions and their derivatives
before forwarding them to the optimizer.

\paragraph{\textbf{Recording.}}
The optimizers in modOpt are equipped with a \texttt{record} object.
When recording is enabled, all problem metadata, 
input variable values, and the associated results of 
function and derivative evaluations are systematically 
logged into the \texttt{record} object.
Any data generated by the optimization algorithm during 
the iterations is similarly logged.
Records are stored as HDF5 files, facilitating further analysis
and visualization post-optimization.
modOpt provides post-processing tools to load
and visualize recorded data.
Since HDF5 files are incompatible with text editors, 
users may set the \texttt{readable\_outputs} option during 
optimizer instantiation to export optimizer-generated data 
as plain text files.
For each variable listed in \texttt{readable\_outputs}, a separate file
is created, with rows representing optimizer iterations.

\paragraph{\textbf{Hot-starting.}}
An optimization run may terminate before achieving sufficient
convergence for various reasons, such as reaching the iteration
limit or setting a loose convergence tolerance.
In such instances, users may choose to re-execute the optimization 
with a higher iteration limit or a tighter tolerance 
to obtain a more refined solution.
The hot-starting feature in modOpt leverages the record file 
to efficiently reuse previously computed function and 
derivative values, thereby avoiding unnecessary computational costs.
This is particularly beneficial when the optimization
functions and their derivatives are computationally expensive.

One key benefit of hot-starting, as opposed to simply restarting
from the previous solution, is that quasi-Newton methods maintain 
the same Hessian approximations from the prior run during a hot-start, 
whereas restarting initializes the Hessian as an identity matrix, 
potentially increasing the number of additional iterations required for convergence.
Furthermore, restarting resets the Lagrange multipliers, which may necessitate
additional iterations, as the convergence of optimization variables
often follows the convergence of Lagrange multipliers.

\paragraph{\textbf{Live-visualization.}}
Every optimizer in modOpt integrates a \texttt{Visualizer} object.
By setting the \texttt{visualize} option to a non-empty list 
of scalar variables when instantiating an optimizer, 
users can activate real-time visualization of those variables. 
This functionality allows users to track the evolution of critical variables
and is particularly valuable during prolonged optimization runs.
The data available for visualization include the optimization
variables, objective, constraints, derivatives, and other 
metrics provided by the optimizer.

\paragraph{\textbf{Benchmarking.}}
modOpt incorporates several utilities designed to facilitate optimizer development.
It integrates with the CUTEst \cite{gould2015cutest} test problem collection
via the PyCUTEst \cite{fowkes2022pycutest} interface.
CUTEst is a comprehensive test suite for nonlinear optimization, comprising over
1500 test problems that span small- and large-scale problems.
Access to these test problems within modOpt enables developers to evaluate
new ideas and benchmark new algorithms against established 
optimizers in the library.
In addition to the test-suite, modOpt includes built-in, scalable examples 
that can serve as benchmarks.
It also provides functions for generating performance profiles
to study the comparative performance of algorithms.
Example scripts demonstrating these capabilities are provided in the
documentation.

\paragraph{\textbf{Testing and documentation.}}
The modOpt repository is hosted on Github, with integrated
continuous integration/continuous delivery (CI/CD) workflows
managed by GitHub Actions to ensure code consistency, 
reliability and maintainability.
The package includes comprehensive tests covering 
all critical classes and interfaces,
with pytest serving as the primary testing framework.
Commits and pull requests to the main branch of the repository
trigger automatic testing.

modOpt is released as an open-source project under the 
GNU Lesser General Public License v3.0
and welcomes contributions from developers.
Documentation for the library is generated using Sphinx 
and hosted on Read the Docs, providing detailed and up-to-date 
information on features, usage, and API references.
The documentation is automatically built and deployed 
following updates to the main branch of the repository,
ensuring both accuracy and accessibility.
By adhering to the latest PyPA standards for Python packaging, 
modOpt maintains high software quality and usability.

    \section{Optimization algorithms in modOpt}
\label{sec:opt}
This section presents an overview of the optimization algorithms
currently available in modOpt.
Section \ref{sec:opt_imp} reviews the algorithms that are natively implemented 
in modOpt, while Section \ref{sec:opt_int} introduces
the algorithms that are integrated with modOpt.

\subsection{Optimizers implemented in modOpt}
\label{sec:opt_imp}
A variety of optimization algorithms are implemented in modOpt.
These algorithms are written entirely in Python and are intended as references
for students and developers.
Their implementation in Python provides unrestricted access to all 
intermediate optimization data, offering a level of transparency 
that contrasts with the black-box nature of many algorithms that are 
interfaced with modOpt.
This subsection provides a concise review of these algorithms, 
which may be skipped by readers already familiar with optimization techniques.
For more detailed information on any of the algorithms discussed below, 
we direct interested readers to the books \cite{martins2021engineering, nocedal1999numerical}.
\paragraph{\textbf{Steepest descent.}}
Also known as the gradient-descent, this method for unconstrained optimization 
computes the direction toward the next iterate $x_{k+1}$ 
based on the direction of steepest-descent.
The variable update follows 
\begin{equation}
    x_{k+1} = x_k - \alpha_k \nabla f(x_k),
    \label{eq:gradient-descent}
\end{equation} 
where $\nabla f(x_k)$ is the gradient of the objective function $f$
and the step size $\alpha_k > 0$ is computed by a line search algorithm.
\paragraph{\textbf{Newton’s method.}}
Newton's method for unconstrained optimization applies the Newton's method for
root-finding to solve $\nabla f = 0$.
The variable update for Newton's method is given by
\begin{equation}
    x_{k+1} = x_k - \alpha_k [\nabla^2 f(x_k)]^{-1} \nabla f(x_k),
    \label{eq:Newton}
\end{equation}
where $\nabla^2 f(x_k)$ represents the Hessian of the objective function $f$.
Newton's method, being a second-order method, can yield an order-of-magnitude faster
solution compared to first-order methods such as gradient-descent.

\paragraph{\textbf{Quasi-Newton methods.}}
In cases where Hessians are unavailable or costly to compute,
various Hessian approximations are employed within Newton's method 
to accelerate optimization convergence.
Quasi-Newton methods are a class of techniques that approximate Hessians at each
iteration ensuring that the curvature of the approximated Hessian 
$\widehat{H}_{k+1}$ along the variable update step $d_k$ closely resembles 
the curvature of the actual Hessian $H_{k+1}$.
This is achieved by enforcing the quasi-Newton condition 
$\widehat{H}_{k+1} d_k = w_k$,
where $d_k = x_{k+1} - x_k$, and $w_k = \nabla f(x_{k+1}) - \nabla f(x_k)$ 
represents the change in the gradient along $d_k$.
The Hessian approximation $\widehat{H}_0$ is typically initialized as an 
identity or diagonal matrix.
Some common quasi-Newton update formulas are given below.
\newline
\textit{Broyden}: 
% A rank-one Hessian update
\begin{equation}
    \widehat{H}_{k+1} = \widehat{H}_k + \frac{1}{d_k^T d_k} (w_k-\widehat{H}_kd_k) d_k^T
    \label{eq:broyden}
\end{equation}
\textit{Symmetric rank-one (SR1)}: 
% A rank-one update that ensures hereditary symmetry
\begin{equation}
    \widehat{H}_{k+1} = \widehat{H}_k 
    + \frac{1}{(w_k-\widehat{H}_kd_k)^T d_k} (w_k-\widehat{H}_kd_k) (w_k-\widehat{H}_kd_k)^T
    \label{eq:sr1}
\end{equation}
\textit{Broyden--Fletcher--Goldfarb--Shanno (BFGS)}: 
% A rank-two update that 
% ensures hereditary symmetry and positive definiteness when $w_k^T d_k > 0$
\begin{equation}
    \widehat{H}_{k+1} = \widehat{H}_k 
    - \frac{1}{d_k^T\widehat{H}_kd_k} \widehat{H}_k d_k d_k^T \widehat{H}_k
    + \frac{1}{w_k^T d_k} w_k w_k^T
    \label{eq:bfgs}
\end{equation}

\paragraph{\textbf{Method of Newton-Lagrange.}}
This method addresses equality-constrained problems by finding
the $(x,\lambda)$ values that satisfy $\nabla L = 0$ using the Newton's method.
Here, $L(x, \lambda)= f(x) - \lambda^T c(x)$ represents the Lagrangian, with 
$\lambda \in \Real^{m}$ being the vector of Lagrange multipliers.
The variable update equation is given by
\begin{equation}
    \begin{bmatrix}
        x_{k+1} \\ \lambda_{k+1} 
    \end{bmatrix}  
    =
    \begin{bmatrix}
        x_k \\ \lambda_k 
    \end{bmatrix}   
    % [x_{k+1}^T, \lambda_{k+1}^T]^T = [x_k^T, \lambda_k^T]^T 
    - \alpha_k [\nabla^2 L(x_k,\lambda_k)]^{-1} \nabla L(x_k,\lambda_k), 
    \label{eq:Newton-Lagrange}
\end{equation}
where
\begin{equation}
    \nabla L(x_k,\lambda_k) = 
    \begin{bmatrix}
        \nabla f(x_k) - J(x_k)^T \lambda_k\\
        -c(x_k)
    \end{bmatrix} , \;
    \text{and} \; 
    \nabla^2 L(x_k,\lambda_k) = 
    \begin{bmatrix}
        H(x_k,\lambda_k) & -J(x_k)^T\\
        -J(x_k) & 0
    \end{bmatrix}.
    \label{eq:Lagrangian derivatives}
\end{equation}
% where
Here, $J(x_k) := \frac{\partial c}{\partial x}\Big\rvert_{x_k} \in \Real^{m \times n}$ 
is the Jacobian of the constraints evaluated at $x_k$, and 
$H(x_k,\lambda_k) := \nabla^2 f(x_k) - \sum_{i=1}^{m} [\lambda_k]_i \nabla^2  c_i(x_k) 
\in \Real^{n \times n}$ 
as the Lagrangian Hessian with respect to $x$.

\paragraph{\textbf{Penalty methods.}}
The \textit{quadratic penalty method} minimizes a sequence of unconstrained optimization problems
with a modified objective that penalizes constraint violations.
For the inequality-constrained problem (\ref{eq:opt}), the penalized objective 
$P_2(x;\rho_k) := f(x) + \frac{1}{2} \rho_k \| c^-(x)\|_2^2$
% \begin{equation}
%     P_2(x;\rho_k) = f(x) + \frac{1}{2} \rho_k || c^-(x)||_2^2
%     \label{eq:l2penalty}
% \end{equation}
is minimized with respect to $x$ for an increasing sequence of positive
penalty parameters $\{\rho_k\}$, where $c^-(x) := -min(c(x),0)$
captures the constraint violations.
The intuition behind this method is that the sequence of unconstrained solutions 
$\{x(\rho_k)\}$ will converge to a local solution $x^*$ of (\ref{eq:opt})
as $\rho_k \to \infty$.

% For sufficiently large values of $\rho$, 
\textit{Exact penalty methods} can solve for a local solution $x^*$ 
using a single unconstrained optimization.
For example, minimizing the $l_1$ penalty function 
$P_1(x;\rho) := f(x) + \rho \| c^-(x)\|_1$ 
or the $l_\infty$ penalty function
$P_{\infty}(x;\rho) := f(x) + \rho \| c^-(x)\|_{\infty}$
for a sufficiently large value of $\rho$ can yield a solution to (\ref{eq:opt}).
For equality-constrained optimization, simply replace $c^-(x)$ with $c(x)$ 
in the penalized objective.

\paragraph{\textbf{Sequential Quadratic Programming (SQP).}}
SQP methods address inequality-constrained problems (\ref{eq:opt}) by
solving a series of Quadratic Programming (QP) subproblems to determine
the search direction $p_k$ towards the next iterate $x_{k+1}$.
Each QP subproblem minimizes a quadratic approximation of the objective
subject to linearized constraints:
\begin{equation}
    \begin{array}{r l}
        \text{minimize}        & f(x_k) + \nabla f(x_k)^T p_k + 
                                \frac{1}{2} p_k^T H(x_k,\lambda_k) p_k           \\
        \text{with respect to} & p_k \\
        \text{subject to}      & c(x_k) + J(x_k)p_k \geq 0.         \\
    \end{array}
    \label{eq:qp}
\end{equation}
The QP solution $(p_k, \widehat \lambda_k)$ is used in a line search with an
appropriate merit function to find the step size $\alpha_k$.
Subsequently, the variables are updated as $x_{k+1} = x_k + \alpha_k p_k$ and
$\lambda_{k+1} = \lambda_k + \alpha_k (\widehat \lambda_k - \lambda_k)$.
% \begin{equation}
%     x_{k+1} = x_k + \alpha_k p_k, \quad \text{and}
%     \lambda_{k+1} = \lambda_k + \alpha_k (\widehat \lambda_k - \lambda_k)
%     \label{eq:sqp-ls}
% \end{equation}
For equality-constrained problems, simply convert the constraints in 
Eq. (\ref{eq:qp}) to equalities.

\paragraph{\textbf{Nelder--Mead method \cite{nelder1965simplex}.}}
Also known as the downhill simplex method or the Nelder--Mead simplex algorithm,
this method is a deterministic, local search algorithm based on heuristics.
It is easy to implement and can handle noisy or discontinuous functions.
The algorithm utilizes a geometric shape called a \textit{simplex}, 
formed by $n+1$ vertices in an $n$-dimensional space,
where each vertex $x^{(j)}$ represents a guess for the optimization variable $x$.

Starting with an initial simplex $\{x^{(0)},x^{(1)}, \dots, x^{(n)} \}$, 
the algorithm iteratively updates the simplex to move it toward
a minimum of the objective function $f$.
The algorithm heuristically adjusts the vertices through geometric operations 
such as reflection, expansion, contraction, or shrinkage, based on the 
relative values of $f$ at each vertex of the current simplex.

\paragraph{\textbf{Particle swarm optimization (PSO).}}
% PSO falls under the broad category of bio-inspired, population-based algorithms.
PSO is a stochastic global search algorithm inspired by swarm intelligence.
The algorithm involves a group of particles, 
each having a position and velocity.
The position represents a candidate solution
while the velocity determines how a particle moves through the solution space.
The goal is for the trajectories of individual particles in the swarm 
to converge to an optimal solution based on the governing heuristics.

Particles start with random positions 
$\{x_0^{(1)},x_0^{(2)},\dots,x_0^{(N)}\}$ and velocities 
$\{v_0^{(1)},v_0^{(2)},\dots,v_0^{(N)}\}$.
New position for the $i$th particle is computed as 
$x_{k+1}^{(i)} = x_k^{(i)} + v_k^{(i)}$, with velocities updated using
$v_{k+1}^{(i)} = wv_k^{(i)} + c_p r_p (x_{best}^{(i)}-x_k^{(i)}) + c_g r_g (x_{best}-x_k^{(i)})$.
% \begin{equation}
%     v_{k+1}^{(i)} = wv_k^{(i)} + c_p r_p (x_{best}^{(i)}-x_k^{(i)}) + c_g r_g (x_{best}-x_k^{(i)}),
%     \label{eq:pso}  
% \end{equation}
Here, $x_{best}^{(i)}$ is the position with the best objective value
found by particle $i$, and $x_{best}$ is the best position found by the swarm.
The constants $w$, $c_p$, and $c_g$ are the inertia weight, 
cognitive parameter, and social parameter, respectively.
Stochasticity is introduced through the random numbers $r_p$ and $r_g$ 
drawn from the interval $[0,1]$.

\paragraph{\textbf{Simulated annealing.}}
Simulated annealing is a stochastic global search algorithm
inspired by metallurgical annealing.
It is more commonly applied to problems with discrete variables
but can also address continuous variables.
At each iteration, the algorithm randomly generates a new point
$x_{new}$ that is sufficiently close to the current iterate $x_k$.
If $f(x_{new}) \leq f(x_k)$, then $x_{k+1}=x_{new}$;
otherwise, $x_{new}$ may be accepted with a probability 
$exp(-\frac{f(x_{new}-f(x_k)}{T_k})$,
where $T_k$ is an artificial temperature parameter.
Higher temperatures favor a global search, while lower temperatures lead to
more focused local searches.
The optimization starts at higher temperatures and slowly cools down
according to some cooling schedule, e.g., $T_k=T_0(1-k/k_{max})$.

\subsection{Optimizers interfaced with modOpt}
\label{sec:opt_int}
Unlike other libraries that are often restricted to specific classes 
of optimization problems, modOpt supports a broad spectrum of problem types,
including nonlinear programming, quadratic programming, and convex programming.
The library offers algorithms tailored for unconstrained, bound-constrained,
equality-constrained, and inequality-constrained problems.

A key advantage of modOpt is its seamless integration with well-maintained,
precompiled packages that are easily installable, eliminating the need for users 
to compile low-level code. 
This feature makes the optimizers more accessible, particularly for students 
and novice programmers, while also allowing users to benefit from 
continuous updates to the original algorithms.
In contrast, other libraries such as pyOpt and pyOptSparse include source code 
that requires user compilation, potentially creating a barrier for 
less experienced users.

This subsection provides a brief overview of each optimizer 
integrated with modOpt.

\paragraph{\textbf{SLSQP \cite{kraft1988software}.}}
Sequential Least SQuares Programming (SLSQP) is an SQP algorithm
designed for solving general nonlinear programming problems.
It builds upon the work of Wilson, Han, and Powell, 
incorporating a BFGS update to estimate the Hessian of the Lagrangian.
SLSQP requires the first derivatives of the objective and constraints 
and is suitable for solving problems with a few hundred variables and constraints.
SLSQP uses an $l_1$ penalty function in the line search to determine the step length.
The search direction is obtained by solving a QP subproblem
after transforming it into an equivalent constrained linear 
least squares problem \cite{schittkowski1982nonlinear}.
The equivalent problem is solved with the linear least squares solver
of Lawson and Hanson \cite{lawson1995solving}.

The original algorithm is implemented in Fortran.
However, precompiled binaries with Python bindings are available through 
the SciPy and PySLSQP \cite{joshy2024pyslsqp} libraries.
modOpt provides interfaces to both libraries.
Notably, PySLSQP offers more extensive access to internal optimization data and 
additional utilities compared to SciPy.

\paragraph{\textbf{SNOPT \cite{gill2005snopt}.}}
The Sparse Nonlinear OPTimizer (SNOPT) is an active-set SQP algorithm designed 
to address large-scale nonlinear programming problems with smooth objective and 
constraint functions. 
It employs a limited-memory BFGS algorithm to approximate the Lagrangian Hessian 
and a smooth augmented Lagrangian merit function for the line search. 
Although SNOPT is capable of solving problems with tens of thousands of variables 
and constraints, it is most effective for problems with up to approximately 
2000 degrees of freedom. 
SNOPT is particularly beneficial for problems whose functions and derivatives are 
costly to compute.
It can also take advantage of the sparsity in constraint Jacobians. 
SNOPT is a commercial software written in Fortran, 
with interfaces to several programming and modeling languages.

\paragraph{\textbf{IPOPT \cite{wachter2006implementation}.}}
The Interior Point OPTimizer (IPOPT) is an algorithm for large-scale 
nonlinear programming.
It is a primal-dual interior point method that solves a series of 
equality-constrained barrier problems to approach the solution of the 
original problem.
Interior point methods can typically handle more variables than active-set
SQP methods since it avoids the combinatorial problem of identifying the
active constraints at a solution.
Accordingly, IPOPT has been successfully applied to problems 
with millions of variables.
IPOPT can exploit exact second derivatives,
but can also approximate them using limited-memory BFGS or SR1 formulas
when Hessians are unavailable.
IPOPT employs a filter-based line search method with second-order corrections
to guarantee global convergence.
IPOPT is implemented in C++, and modOpt interfaces with its precompiled version 
available through CasADi \cite{andersson2019casadi}.

\paragraph{\textbf{trust-constr \cite{virtanen2020scipy}.}}
The trust-constr algorithm is a trust-region-based large-scale 
nonlinear programming method.
When Hessians are not provided, it uses the BFGS method to approximate them.
For equality-constrained problems, the algorithm employs the Byrd-Omojokun 
trust-region SQP method \cite{lalee1998implementation, nocedal1999numerical},
which solves the intermediate QP subproblems by decomposing them into 
two smaller unconstrained trust-region subproblems.
For inequality-constrained problems, trust-constr switches to a 
trust-region interior point algorithm \cite{byrd1999knitro}, where the 
equality-constrained barrier subproblems are solved approximately using the 
trust-region SQP method.
The numerical implementation follows the techniques outlined by 
Conn et al. \cite{conn2000trust} and is fully realized in SciPy, only using Python.

\paragraph{\textbf{L-BFGS-B \cite{byrd1995limited}.}}
L-BFGS-B is one of the most efficient algorithms for solving large-scale 
bound-constrained and unconstrained optimization problems.
It uses a limited memory BFGS algorithm \cite{byrd1994representations} 
to approximate the Hessian of the objective function
and employs a gradient projection method to identify the set of active bounds 
during each iteration.
Although the original implementation \cite{zhu1997algorithm} is in Fortran, 
modOpt uses the precompiled version available through SciPy.

\paragraph{\textbf{BFGS \cite{nocedal1999numerical}.}}
As discussed in Sec. \ref{sec:opt_imp}, the BFGS algorithm 
is a quasi-Newton method for solving unconstrained optimization problems.
The performant version of the BFGS algorithm in modOpt leverages the 
BFGS method from the SciPy library.
It is entirely written in Python and uses inverse BFGS updates.

\paragraph{\textbf{Nelder-Mead \cite{nelder1965simplex}.}}
The performant version of the Nelder-Mead algorithm in modOpt 
utilizes the Python implementation from SciPy and is designed to 
solve bound-constrained problems.
This gradient-free, direct search method implements an adaptive variant 
\cite{gao2012implementing} of the standard Nelder-Mead algorithm 
(discussed earlier in Sec. \ref{sec:opt_imp}),
where the expansion, contraction, and shrinkage parameters are adjusted 
according to the dimension of the problem.
The adaptive approach has been shown to outperform the standard method,
particularly in high-dimensional problems with up to 60 optimization variables. 

\paragraph{\textbf{COBYLA \cite{powell1994direct}.}}
Constrained Optimization BY Linear Approximation (COBYLA) is a gradient-free 
algorithm designed for low-dimensional nonlinear programming problems 
with only inequality constraints.
It operates by approximating the objective and constraint functions 
through linear interpolation at the $n+1$ vertices of a simplex.
In each iteration, COBYLA solves a linear programming problem within 
a spherical trust region, updating a simplex vertex if the new point
calculated is the best found so far, according to a merit function.
In some iterations, a vertex may also be updated to improve 
the linear models of the objective and constraints.
COBYLA in modOpt uses the precompiled version of the original
Fortran implementation available from SciPy.

\paragraph{\textbf{COBYQA \cite{ragonneau2022model}.}}
COBYQA (Constrained Optimization BY Quadratic Approximations) is a 
derivative-free, trust-region SQP method for solving general
nonlinear programming problems.
It constructs quadratic trust-region models using 
underdetermined interpolation, based on the derivative-free 
symmetric Broyden update proposed by Powell \cite{powell2004least}.
As a successor to COBYLA, COBYQA has demonstrated consistently 
superior performance over COBYLA in solving problems with up 
to 50 variables \cite{ragonneau2022model}.
COBYQA is available as an open-source package implemented in Python,
with some of its solvers optimized using Cython.

\paragraph{\textbf{QP solvers \cite{qpsolvers2024}.}}
Quadratic programming refers to techniques used for solving 
optimization problems characterized by a quadratic objective 
function and linear constraints. 
modOpt integrates with qpsolvers, a Python library that serves 
as a unified interface for more than 15 convex QP solvers. 
The qpsolvers library features algorithms such as active set, 
interior-point, and augmented Lagrangian methods,
and is capable of addressing both sparse and dense formulations.
Additionally, it includes several benchmarks for performance 
comparison among QP solvers.

\paragraph{\textbf{CVXOPT \cite{andersen2020cvxopt}.}}
CVXOPT is an open-source Python library specifically designed 
for solving convex optimization problems.
It includes solvers tailored for linear and quadratic cone programming, 
second-order cone programming, semidefinite programming, and 
geometric programming, among other convex programming methods.
modOpt interfaces with CVXOPT’s general nonlinear convex optimization solver,
supporting problems with objective and constraint functions
that are convex and twice differentiable.
This implies that all equality constraints must be linear.

    \section{Numerical studies}
    \label{sec:egs}
    This section presents various numerical studies
    conducted using the modOpt library.
    In Section \ref{sec:small_scale}, we evaluate different educational and high-performance
    algorithms in modOpt using three simple analytical problems.
    Section \ref{sec:scalable} examines the scalability of optimizers 
    as we apply them to two distinct versions of the multidimensional Rosenbrock problem.
    Section \ref{sec:eng_apps} investigates the suitability of the modeling
    languages and optimizers in practical applications, utilizing 
    two simplified engineering problems as case studies.
    In Section \ref{sec:cutest_bm}, we benchmark optimizers on a set of 
    unconstrained problems from the CUTEst test-suite.
    
    We conduct all studies, with the exception of those presented in
    Section \ref{sec:eng_apps}, on a Macbook Pro with a 
    quad-core Intel Core i5 processor (2.4 GHz) and 8 GB of RAM.
    Due to the memory-intensive nature of the problems in 
    Section \ref{sec:eng_apps}, we utilize an Ubuntu desktop system 
    with a 16-core Intel Core i7 processor (3.4 GHz) and 112 GB of RAM
    for those studies.
    For the purposes of a fair comparison, all optimizers were configured 
    to use their default parameter settings.
    
    \subsection{Small-scale analytical problems}
    \label{sec:small_scale}
    We start by applying optimization algorithms on
    three carefully selected unconstrained problems.
    Among the educational algorithms, we consider steepest descent, 
    Newton, and BFGS algorithms, with and without the use of line searches.
    Since the steepest descent method without a line search failed to converge 
    on all problems, we exclude it from this study.
    For the performant algorithms, we evaluate SNOPT, PySLSQP,
    BFGS, L-BFGS-B, COBYLA, COBYQA, Nelder--Mead, and trust-constr.
    As the instructional algorithms are all gradient-based,
    we use the gradient norm as the optimality measure in the convergence analysis.
    When comparing performant algorithms, we utilize the objective function value
    as the optimality measure, since some algorithms are gradient-free.
    
    \paragraph{\textbf{Quartic function.}}
    We first consider the following two-dimensional quartic function
    in Eq. (\ref{eq:quartic}) to study the accuracy and convergence rate 
    of the algorithms.
    This function has a very low curvature in the vicinity of the solution $(0,0)$,
    which poses a challenge for some methods.
    
    \begin{equation}
        \begin{array}{r l}
            \min \limits_{x_1,x_2}        & f(x) = x_1^4 + x_2^4            \\
        \end{array}
        \label{eq:quartic}
    \end{equation}
    
    \begin{figure}[ht]
        \centering
        \includegraphics[width=0.45\linewidth]{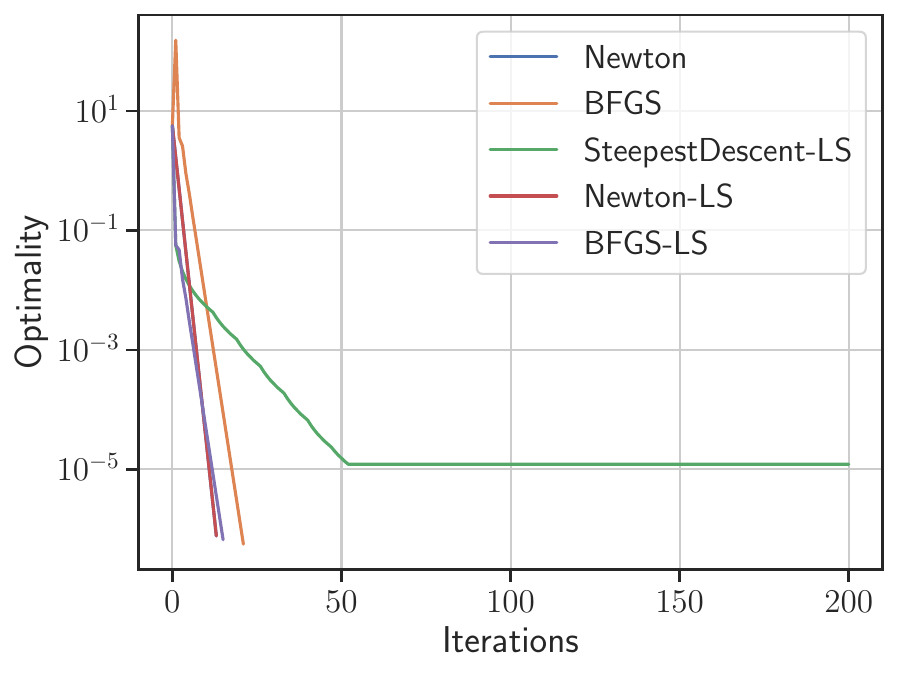}
        \includegraphics[width=0.45\linewidth]{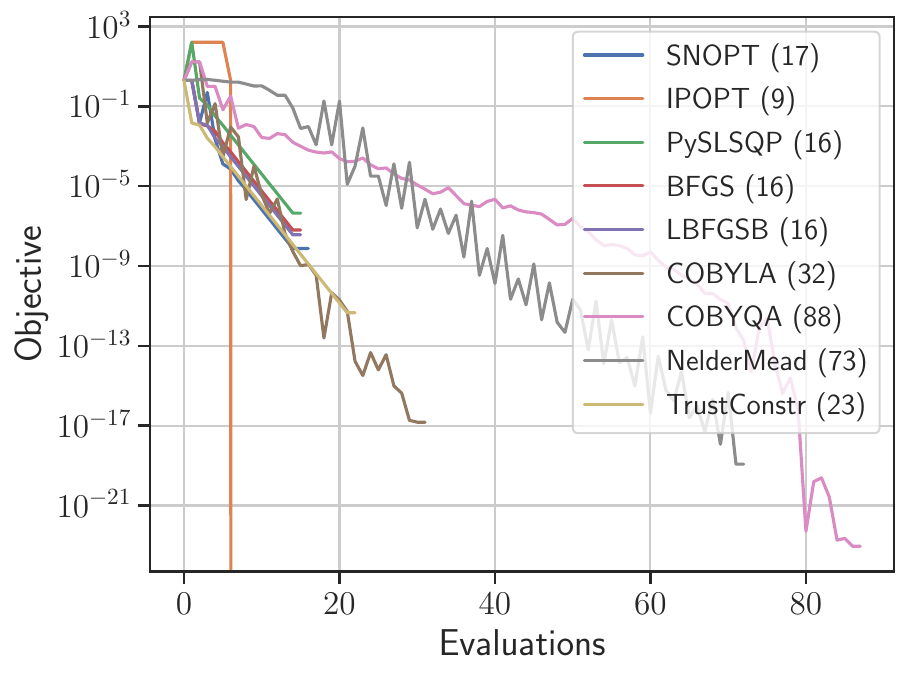}
        \caption{\textbf{Convergence comparison of different algorithms using the quartic function.} 
                Every algorithm starts from the initial guess $(1, 1)$.
                The total number of objective function evaluations required 
                for convergence is indicated in parentheses next to the name 
                of each algorithm in the legend.
            % \normalfont{}
        }
        \label{fig:quartic}
    \end{figure}
    
    Figure \ref{fig:quartic} illustrates the convergence behavior of various methods
    applied to the quartic function problem.
    Despite incorporating a line search, the steepest descent method
    demonstrates the slowest convergence and it fails to 
    reduce the optimality gap below $1e-5$.
    This is due to the lack of curvature information and the small gradients
    near the solution, which cause steepest descent to take very small steps.
    Newton's method, both with or without a line search,
    shows the same convergence behavior, since the line search consistently
    selects a unit step size for this problem.
    Newton's method also achieves the fastest convergence in this problem.
    The BFGS algorithm exhibits a similar convergence rate and final optimality 
    as the Newton's method. 
    When combined with a line search, BFGS shows improved convergence, 
    nearly identical to that of Newton's method.
    
    Among the performant algorithms, IPOPT finds the exact solution,
    while other gradient-based methods and COBYLA converge at
    comparable rates.
    The other gradient-free methods, COBYQA and Nelder--Mead,
    takes about three times the number of function evaluations
    to achieve similar optimality as other methods.
    However, this does not result in a significantly higher computational cost, 
    as gradient-based methods evaluate both the objective function and 
    its gradient at each iteration.
    Overall, IPOPT and COBYLA emerge as the most efficient
    algorithms in this study, considering only the number of function and 
    gradient evaluations.

    \paragraph{\textbf{Rosenbrock problem.}}
    The Rosenbrock function (\ref{eq:rosenbrock}) is a well-known 
    benchmark for testing nonlinear optimization algorithms.
    It is a non-convex function characterized by a long, narrow valley with steep curvatures surrounding the solution.
    This problem is particularly challenging as algorithms must 
    navigate regions of high curvature to locate the global minimum.
    Although many methods approach the valley, they often struggle to converge 
    to the global minimum at $x=(1,1)$, where the objective value is zero.
    
    \begin{equation}
        \begin{array}{r l}
            \min \limits_{x_1, x_2}        & f(x) = (1-x_1)^2  + 100(x_2 - x_1^2)^2    \\
        \end{array}
        \label{eq:rosenbrock}
    \end{equation}
    
    \begin{figure}[ht]
        \centering
        \includegraphics[width=0.45\linewidth]{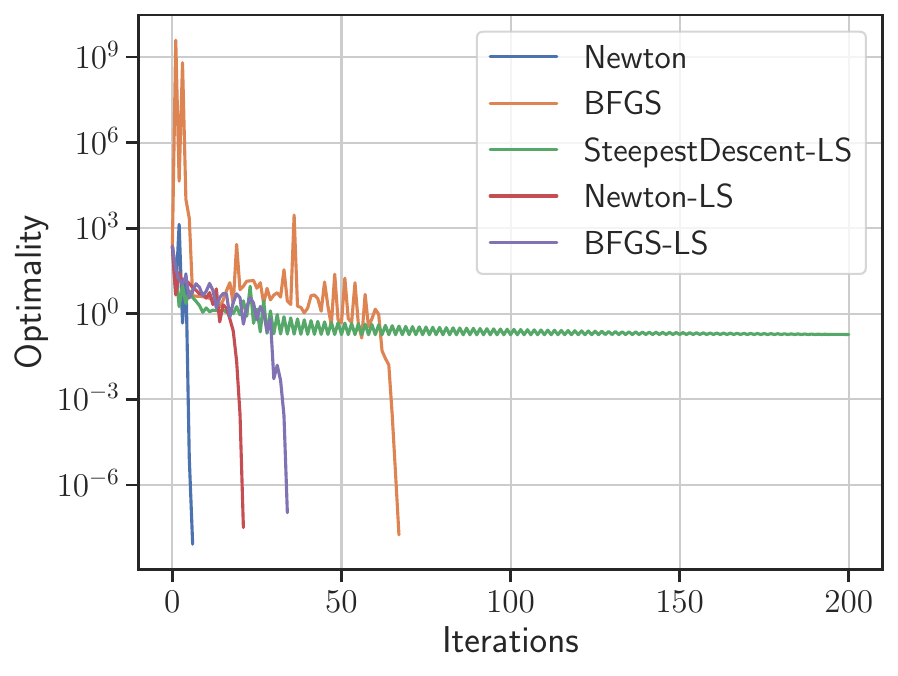}
        \includegraphics[width=0.45\linewidth]{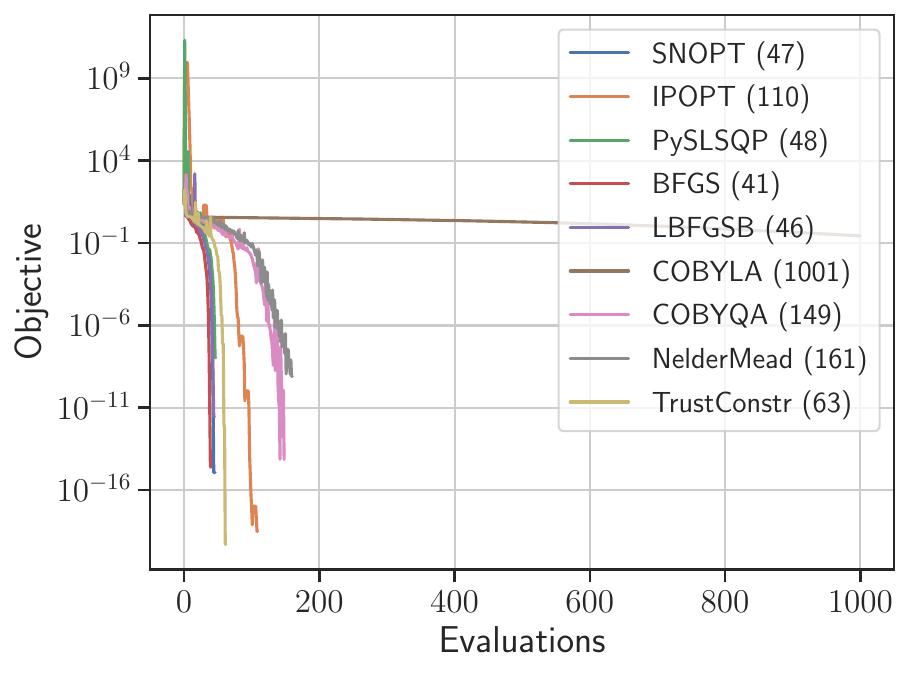}
        \caption{\textbf{Convergence comparison of different algorithms using the Rosenbrock function.}
                        Every algorithm starts from the initial guess $(-1.2, 1)$.
            % \normalfont{}
        }
        \label{fig:rosenbrock}
    \end{figure}
    
    Figure \ref{fig:rosenbrock} indicates that all the 
    standard algorithms converge differently for this problem.
    The steepest descent method fails to identify the solution,
    oscillating around a gradient norm of $0.1$ without further progress.
    The BFGS method with a line search converges twice
    as fast as its counterpart without a line search.
    Conversely, Newton's method exhibits slower convergence when 
    paired with a line search.
    This can be attributed to the fact that Newton's method with
    a line search converges along a different path in the optimization space. 
    Although some iterations of the pure Newton method show an increase in
    optimality, this does not lead to divergence. 
    However, this behavior appears to be incidental and specific to this 
    particular problem.
    For assured convergence, Newton's method typically requires 
    globalization strategies, such as a line search or a trust region method.
    This study highlights the importance of curvature information and
    globalization techniques in optimization.
    
    Among the performant algorithms, all gradient-based methods,
    with the exception of IPOPT, converge within 50 iterations.
    COBYQA, Nelder-Mead, and IPOPT require approximately 150 iterations 
    to reach the solution, while COBYLA fails to converge.
    The failure of COBYLA can be attributed to its reliance on
    linear approximations of the objective function, which proves insufficient for
    capturing the high variations in curvature present in the Rosenbrock function.
    
    \paragraph{\textbf{Bean function.}}
    The bean function (\ref{eq:bean}) was introduced in 
    \cite{martins2021engineering} as a milder variant of 
    the Rosenbrock function, characterized by smoother curvature variations.
    It serves as a benchmark to assess the performance of algorithms on
    problems with moderate curvatures.
    The global minimum of this function is located at $x = (1.21314, 0.82414)$,
    with an objective value of $0.09194$.
    
    \begin{equation}
        \begin{array}{r l}
            \min \limits_{x_1, x_2}        & f(x) = (1 - x_1^2)^2 + (1 - x_2^2)^2 + 
            \frac{1}{2}(2x_2-x_1^2)^2             \\
        \end{array}
        \label{eq:bean}
    \end{equation}
    
\begin{figure}[ht]
        \centering
        \includegraphics[width=0.45\linewidth]{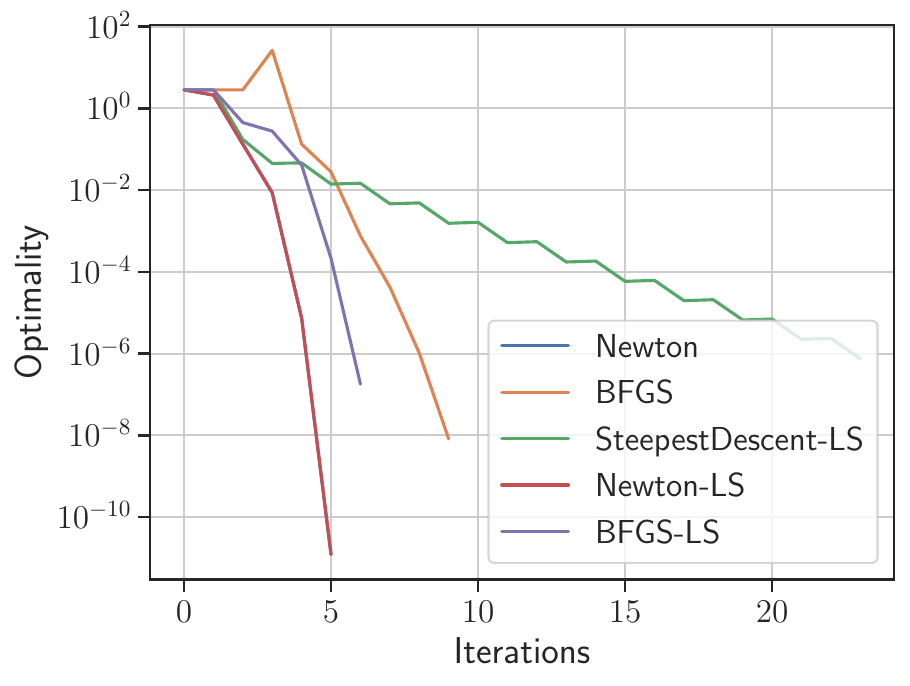}
        \includegraphics[width=0.45\linewidth]{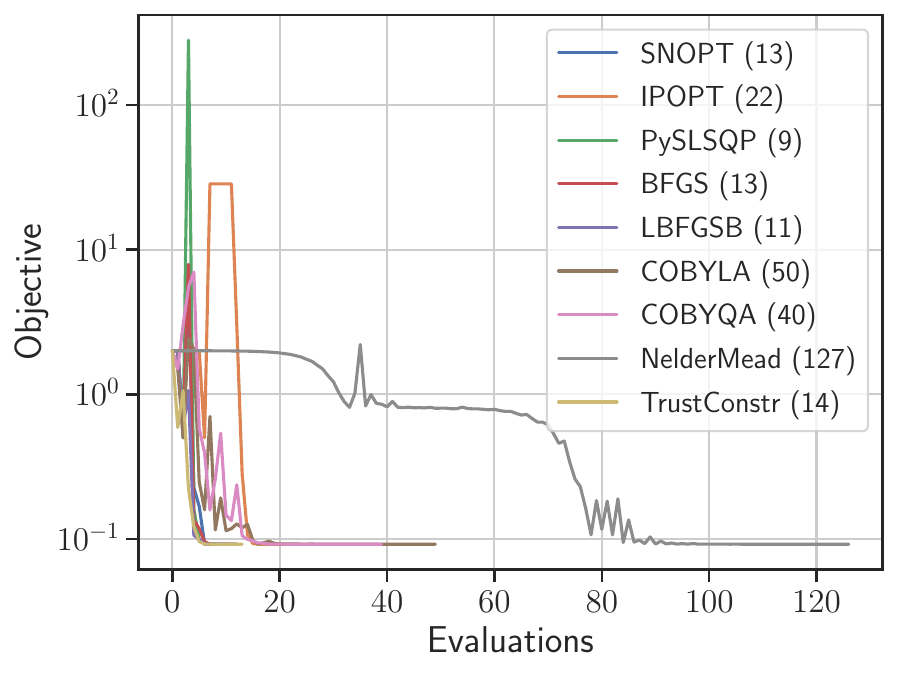}
        \caption{\textbf{Convergence comparison of different algorithms using the bean function.}
                         Every algorithm starts from the initial guess $(0, 0)$.
            % \normalfont{}
        }
        \label{fig:bean}
    \end{figure}

    As shown in Fig. \ref{fig:bean}, the convergence behavior of the 
    instructional algorithms on this function is largely consistent 
    with their performance on the quartic problem discussed earlier.
    The key exception is that the steepest descent method successfully converges,
    unlike in the quartic case where it encountered difficulties.
    This suggests that for problems with moderate curvatures, steepest descent
    is capable of converging, although at a significantly slower rate---approximately 
    one order of magnitude slower than the BFGS or Newton methods.
    This result underscores the critical role of curvature information 
    in achieving rapid convergence, even in smooth problems.
    
    Owing to the smoother characteristics of the Bean function, 
    all high-performing methods, except for Nelder–Mead, 
    converge in around 20 iterations. 
    In contrast, Nelder–Mead requires approximately 100 iterations 
    to achieve convergence, further emphasizing the significance of 
    gradient information in relatively smooth problems.
    This is particularly evident as the other gradient-free methods, 
    COBYLA and COBYQA, rely on gradient approximations.
    
    \subsection{Scalable analytical problems}
    \label{sec:scalable}
    
    We now evaluate the performance of optimization algorithms on 
    two scalable variants of the Rosenbrock problem.
    The study was performed by starting with 2 optimization variables 
    and then successively increasing the problem size by a factor of two 
    until an upper limit of 4096 variables for the first problem and 
    512 variables for the second.
    In this study, we also include the second-order IPOPT and trust-constr
    algorithms, which utilize the exact Hessian of the objective function.
    
    \subsubsection*{Multidimensional Rosenbrock - uncoupled}
    The uncoupled Rosenbrock problem (\ref{eq:rosenbrock_u}) is defined
    for an even number of variables $n$.
    The problem features a composite objective derived from summing
    the objectives of $n/2$ distinct two-dimensional Rosenbrock problems.
    It is termed `uncoupled' because it is separable, with its solution 
    coming from those of $n/2$ two-dimensional Rosenbrock problems.
    Consequently, the global minimum is located at 
    $x=(1,1,...,1)$ with an objective value of zero.
    
    \begin{equation}
        \begin{array}{r l}
            \min \limits_{x\in\Real^n}        & f(x) = \sum_{i=1}^{n/2}  100(x_{2i-1} - x_{2i}^2)^2 + (1-x_{2i-1})^2             \\
        \end{array}
        \label{eq:rosenbrock_u}
    \end{equation}
    
    \begin{figure}[ht]
        \centering
        \includegraphics[width=0.45\linewidth]{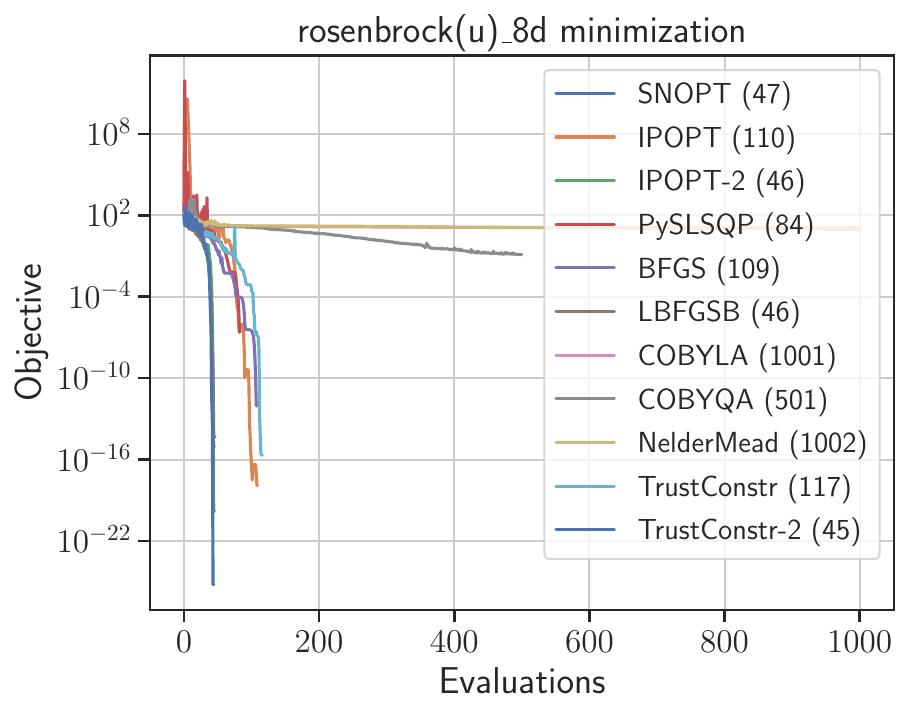}
        \includegraphics[width=0.45\linewidth]{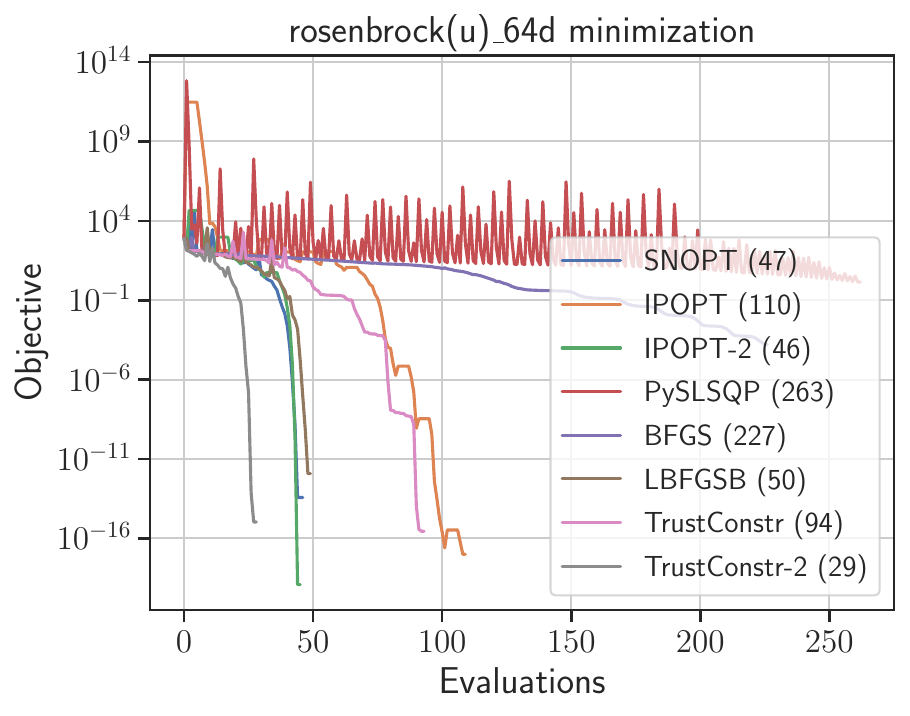}
        \includegraphics[width=0.45\linewidth]{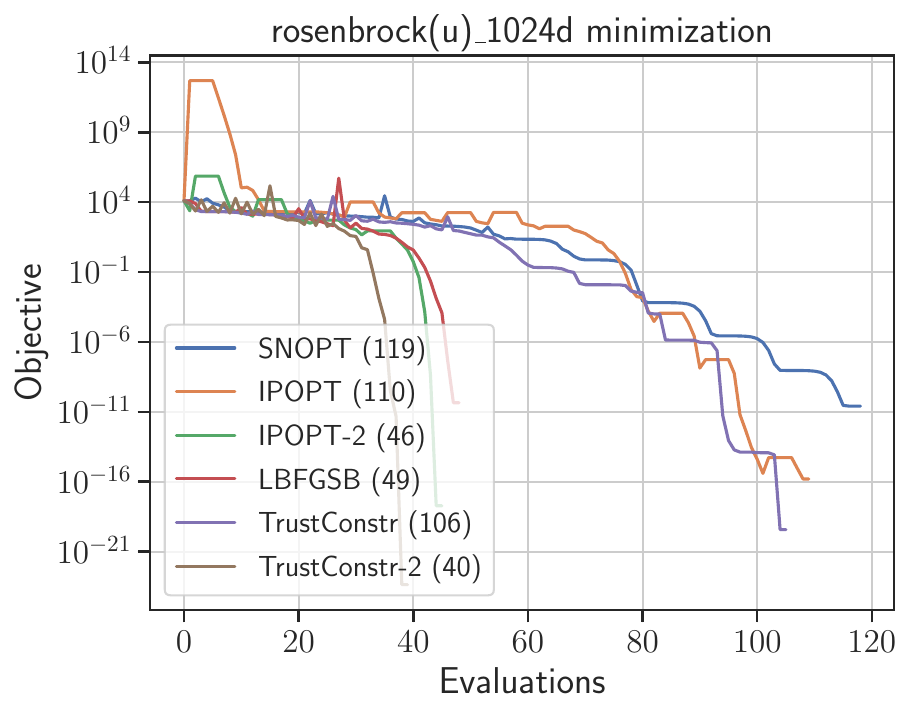}
        \includegraphics[width=0.45\linewidth]{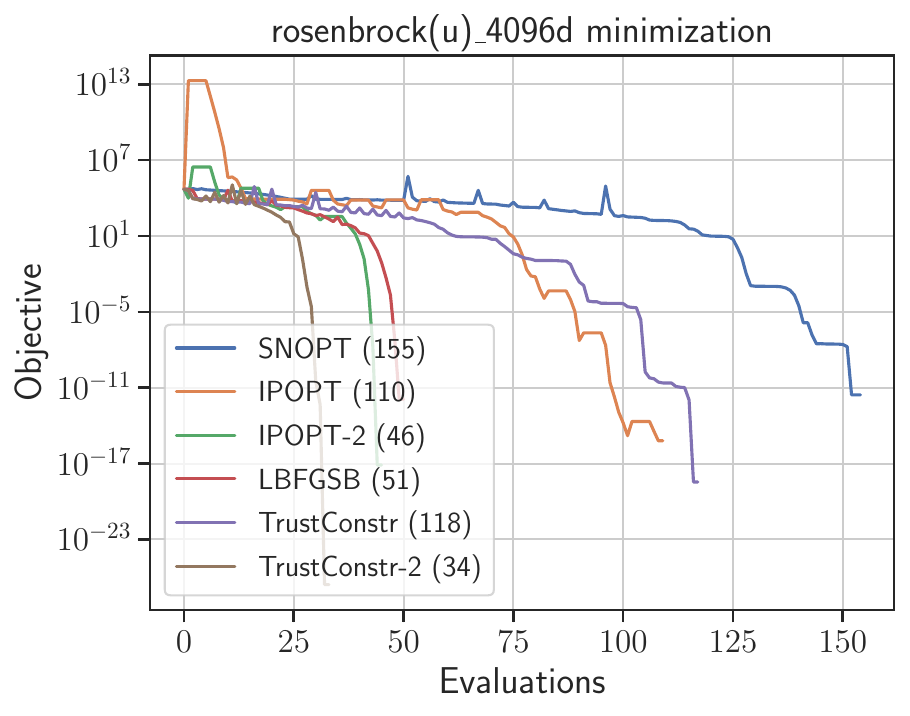}
        \caption{\textbf{Convergence comparison of optimizers using uncoupled Rosenbrock functions.}
        Every algorithm starts from the initial guess $(-1.2, 1, -1.2, 1, ..., -1.2, 1)$.
            % \normalfont{}
        }
        \label{fig:rosenbrock_u}
    \end{figure}
    
    Figure \ref{fig:rosenbrock_u} presents the convergence results for problems with 
    8, 64, 1024, and 4096 variables.
    The gradient-free methods fail to converge for the 8-variable problem,
    while  SNOPT, L-BFGS-B, and the second-order methods exhibit significantly 
    faster convergence compared to other first-order gradient-based approaches.
    As the number of variables increases to 64, PySLSQP and BFGS struggle
    to converge, while other methods maintain similar performance.
    We hypothesize that the spikes observed in the PySLSQP convergence plot 
    result from the derivative-free line search employed in
    the SLSQP algorithm, which makes it difficult to progress through
    regions of high curvature.
    
    For problems with 1024 and 4096 variables,
    L-BFGS-B, despite being a first-order method, converges at a quadratic rate
    comparable to the second-order methods.
    As the number of variables increases, SNOPT requires more iterations to converge.
    Notably, the IPOPT optimizer converges within the same number of evaluations,
    regardless of the problem size.
    This suggests that IPOPT effectively exploits the problem's separable structure, 
    an advantageous feature for large-scale problems where manual identification 
    of separability is impractical.
    
    \subsubsection*{Multidimensional Rosenbrock - coupled}
    The coupled variant of the Rosenbrock problem (\ref{eq:rosenbrock_c}) is defined
    for any number of variables.
    This variant is generally more challenging than the uncoupled version
    due to its non-separability.
    The objective function for this problem is expressed as the sum of two-dimensional 
    Rosenbrock functions applied to each adjacent pair of variables within the 
    optimization variable vector. 
    The global minimum occurs at $x=(1,1,...,1)$ with an objective
    value of zero.
    However, for dimensions greater than four, a local minimum emerges at
    $x=(-1,1,...,1)$ with an objective value of $4$.
    
    \begin{equation}
        \begin{array}{r l}
            \min \limits_{x\in\Real^n}        & f(x) = \sum_{i=1}^{n-1}  100(x_{i+1} - x_i^2)^2 + (1-x_i)^2             \\
        \end{array}
        \label{eq:rosenbrock_c}
    \end{equation}
    
    \begin{figure}[ht]
        \centering
        \includegraphics[width=0.45\linewidth]{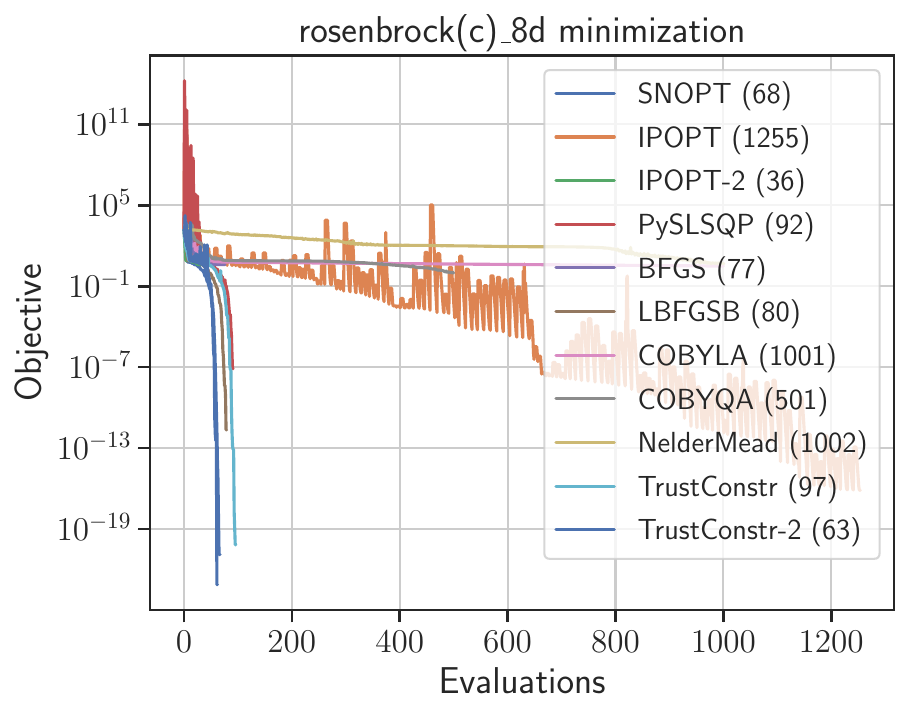}
        \includegraphics[width=0.45\linewidth]{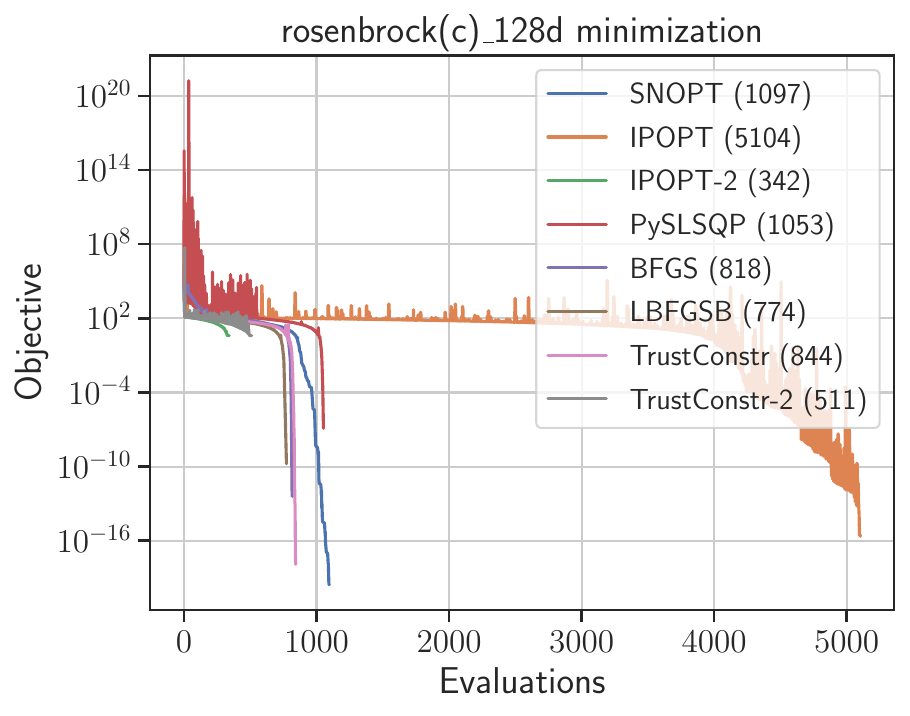}
        \includegraphics[width=0.45\linewidth]{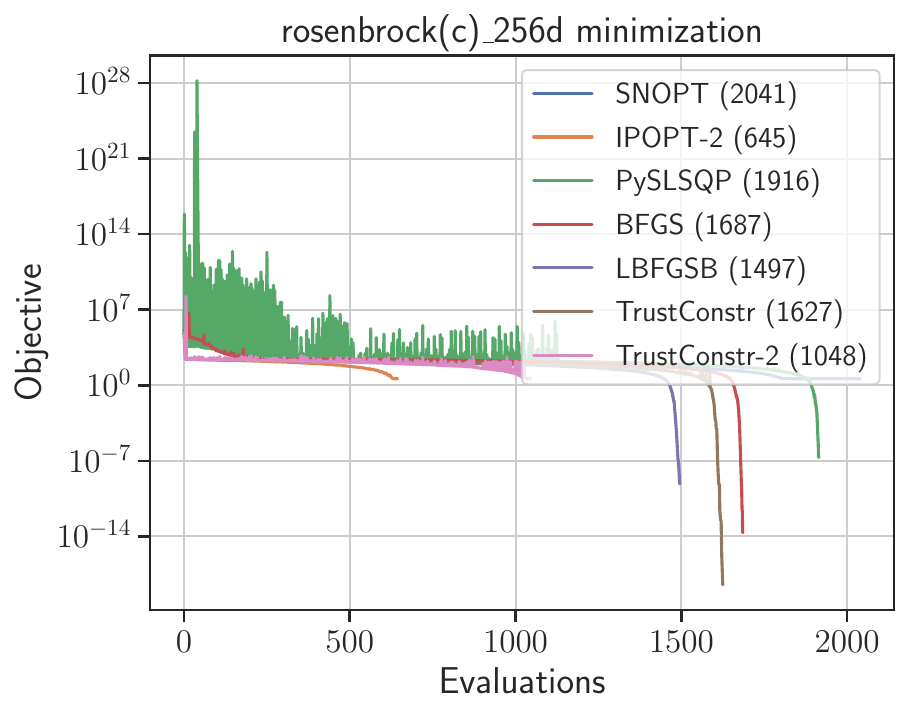}
        \includegraphics[width=0.45\linewidth]{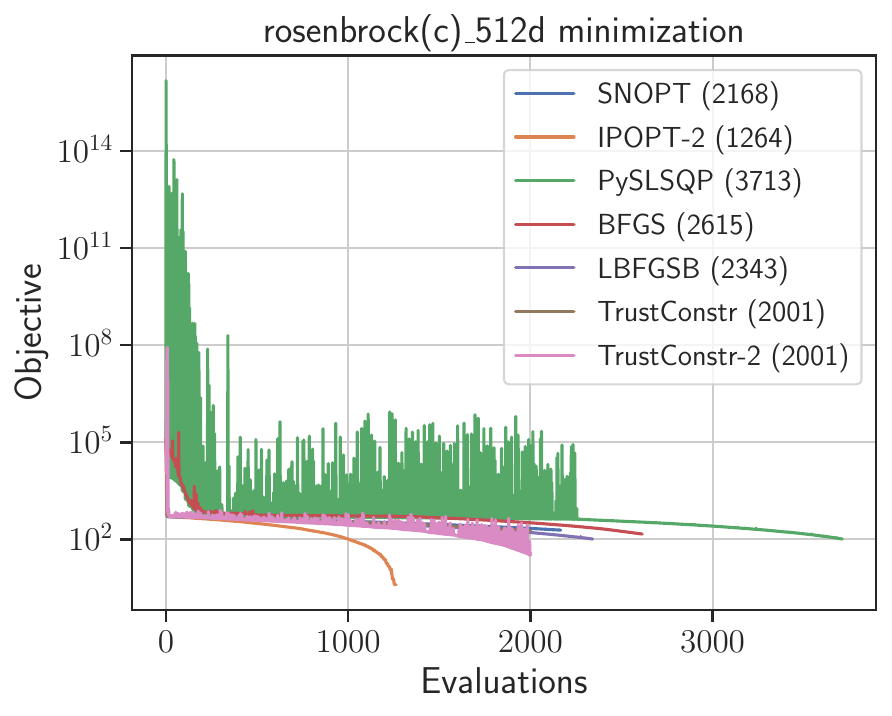}
        \caption{\textbf{Convergence comparison of optimizers using coupled Rosenbrock functions.}
            Every algorithm starts from the initial guess $(-1.2, 1, -1.2, 1, ..., -1.2, 1)$.
            % \normalfont{}
        }
        \label{fig:rosenbrock_c}
    \end{figure}
    
    Figure \ref{fig:rosenbrock_c} illustrates the convergence behavior of
    various optimizers on this problem as the number of variables increases.
    Notably, there are significant differences in optimizer performance
    compared to the uncoupled problem.
    For instance, first-order IPOPT requires over 1000 iterations 
    to converge for the 8-dimensional problem.
    As in the previous case, gradient-free methods fail to find a solution.
    At 128 dimensions, the second-order methods
    begin converging to the local minimum rather than the global one.
    SNOPT ceases to converge beyond 128 dimensions, while other first-order
    methods encounter difficulties at 512 dimensions.
    However, the second-order methods continue to converge even with 512 variables,
    highlighting the critical role of second-derivatives in solving
    high-dimensional problems characterized by steep curvatures.
    
    Although we did not explicitly assess the radius of convergence
    in this study, it is worth noting that second-order IPOPT 
    consistently converged across a wide range of initial guesses 
    for the 512-variable problem, 
    indicating a larger basin of convergence compared to other methods.
    
    Based on the results from this study, we present the following 
    summary and recommendations. 
    For highly nonlinear problems involving 10 or more variables, it is advisable 
    to use gradient-based optimizers, particularly if the models are 
    computationally expensive.
    Among first-order methods for unconstrained problems, 
    L-BFGS-B demonstrated the best performance, making it a suitable initial choice.
    Second-order IPOPT emerged as the most robust optimizer among those tested.
    Additionally, the trust-constr algorithm, both with and without second derivatives, 
    also shows promise as a robust optimizer for unconstrained problems.
    When a problem appears to be separable, IPOPT is recommended, 
    as it can potentially exploit this structure for more efficient convergence.
    
    An unexpected observation was the poor convergence rate of first-order IPOPT, 
    even in low-dimensional instances of the coupled Rosenbrock problem. 
    The cause of this behavior is unclear.
    We anticipate that adjusting the default settings of the optimizer may 
    mitigate this issue.
    For algorithm developers and researchers, we hypothesize that the performance 
    of the SLSQP algorithm could be significantly improved by incorporating a 
    derivative-based line search and a limited-memory BFGS Hessian approximation.
    
    \subsection{Engineering applications}
    \label{sec:eng_apps}
    
    This section presents two scalable engineering case studies
    utilizing the optimizers and modeling options available through modOpt.
    These case studies evaluate the scalability of the algorithms and modeling frameworks 
    by analyzing computational costs in terms of both runtime and memory usage
    as the problem size increases.
    The first problem focuses on optimizing the thickness distribution of a 
    cantilever beam, 
    while the second addresses the optimization of a spacecraft's trajectory and control 
    during its landing phase.
    These problems were purposely chosen:
    the first involves implicitly calculated state variables, 
    allowing us to assess model scaling when implicit differentiation is employed, 
    while the second consists only of explicit equations, serving as a benchmark for 
    models where derivatives are computed through straightforward chain rule application.

    Both problems involve equality and inequality constraints, 
    requiring algorithms capable of addressing 
    nonlinear optimization problems with general constraints. 
    Accordingly, our studies focus on the performance of PySLSQP, SNOPT, 
    trust-constr, and IPOPT optimizers. 
    We also examine both the first- and second-order capabilities of 
    trust-constr and IPOPT. 
    The models are developed in modOpt, CasADi, CSDL, Jax, and OpenMDAO, 
    with Jax and CasADi supporting both first- and second-order formulations.
    Models implemented natively within modOpt employ finite-difference approximations 
    for derivatives, primarily for ease of implementation, 
    while the other frameworks utilize analytical derivatives.
    
    Accurately measuring memory usage in Python programs that incorporate extensions 
    from languages such as C++ or Fortran presents certain challenges. 
    Additionally, an optimization algorithm may invoke different branches of its logic 
    as the problem size changes, 
    making it challenging to observe consistent trends in memory costs 
    during scalability studies. 
    Despite these limitations, we present these preliminary results 
    as a guideline for the development and optimization of models in various 
    engineering applications.
    
    \subsubsection*{Cantilever beam thickness optimization}
    In this problem, we seek to optimize the thickness distribution of a 
    cantilever beam to minimize its compliance under a concentrated tip load.
    The beam is modeled with a rectangular cross-section and is governed by
    Euler-Bernoulli beam theory.
    We impose an equality constraint on the total volume of the beam.
    The optimization problem is formulated as
    \begin{equation}
        \begin{array}{r l}
            \text{minimize}        & F^T d                 \\
            % \min \limits_{h}        & F^T d                 \\
            \text{with respect to} & h \geq 0 \\
            \text{subject to}      & \frac{bL}{n} \sum_{i=1}^{n} h_i = V_0, \\
        \end{array}
        \quad \text{with} \quad K(h) d - F = 0,
        \label{eq:cantilever}
    \end{equation}
    where $F$ is the force vector, $d$ is the displacement vector, 
    $n$ is the number of beam elements,
    $h \in \Real^n$ denotes the beam's thickness distribution,
    $V_0$ is the prescribed total volume, $L$ is the length of the beam, 
    $b$ is the breadth of the beam's cross section,
    and the stiffness matrix $K(h)$ is a function of the thickness distribution.
    The constants employed in this study are:
    $L=1.0 \; m$, $b=0.1 \; m$, $V_0=0.01 \; m^3$, and $F=(0,0,...,0,-1,0) \; N$.
    The modulus of elasticity $E$ is set to $1.0 \; N/m^2$.

    \begin{figure}[ht]
        \centering
        \includegraphics[width=0.45\linewidth]{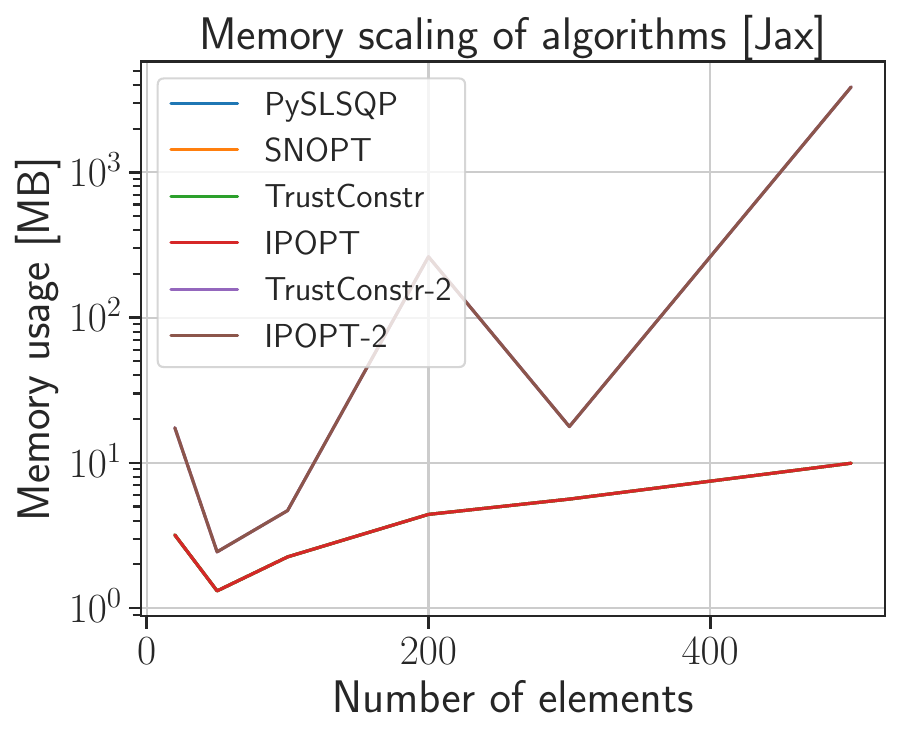}
        \includegraphics[width=0.45\linewidth]{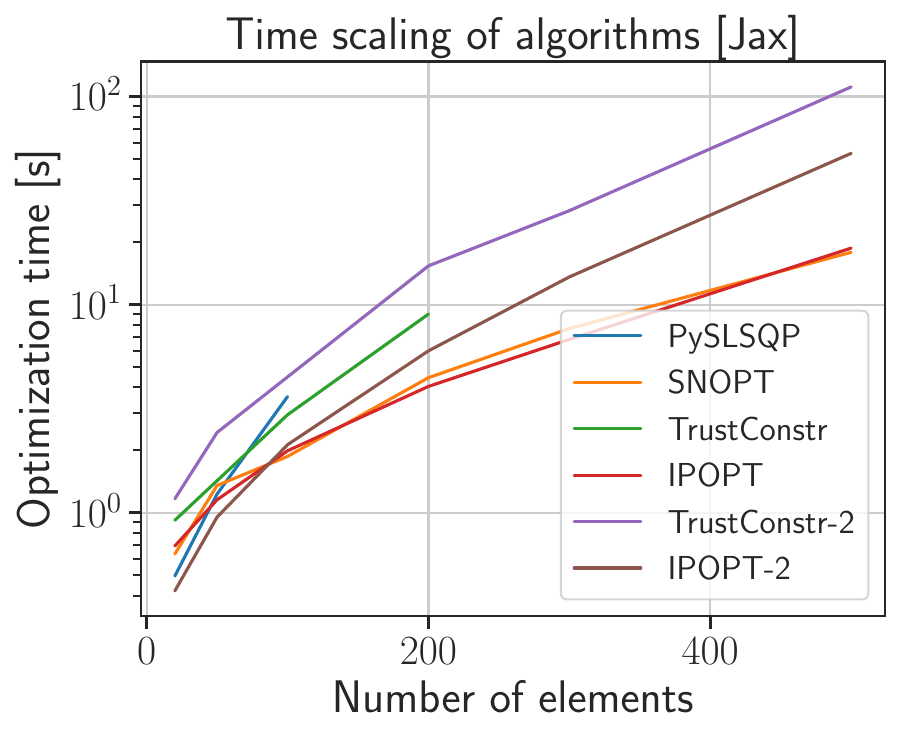}
        \caption{\textbf{Memory usage and solve time comparison of various optimizers in a 
            scaling study using the cantilever beam problem.}
            The beam models are implemented in Jax.
            % \normalfont{}
        }
        \label{fig:cantilever_opts}
    \end{figure}

    To assess the scalability of different optimization algorithms, 
    we compare their time and memory usage as the number of beam elements 
    increases in the sequence: 20, 50, 100, 200, 300, and 500. 
    Figure \ref{fig:cantilever_opts} shows the memory and time scaling 
    for all methods using beam models implemented in Jax.
    Notably, not all optimizers converge as the problem dimension increases.
    PySLSQP stops converging at 100 beam elements while the first-order trust-constr 
    method fails at 200 elements.
    
    In terms of memory usage, first-order methods exhibit approximately linear scaling,
    while second-order methods scale between quadratic and cubic rates.
    Among all tested methods, SNOPT and first-order IPOPT outperform the others 
    in terms of convergence times.
    Despite the reduction in the number of model evaluations achieved by 
    second-order methods, the additional cost of computing Hessians 
    for this problem (involving implicit state variables and loops 
    that scale with the number of beam elements) results in longer 
    convergence times for these methods compared to their first-order counterparts.

    \begin{figure}[ht]
        \centering
        \includegraphics[width=0.45\linewidth]{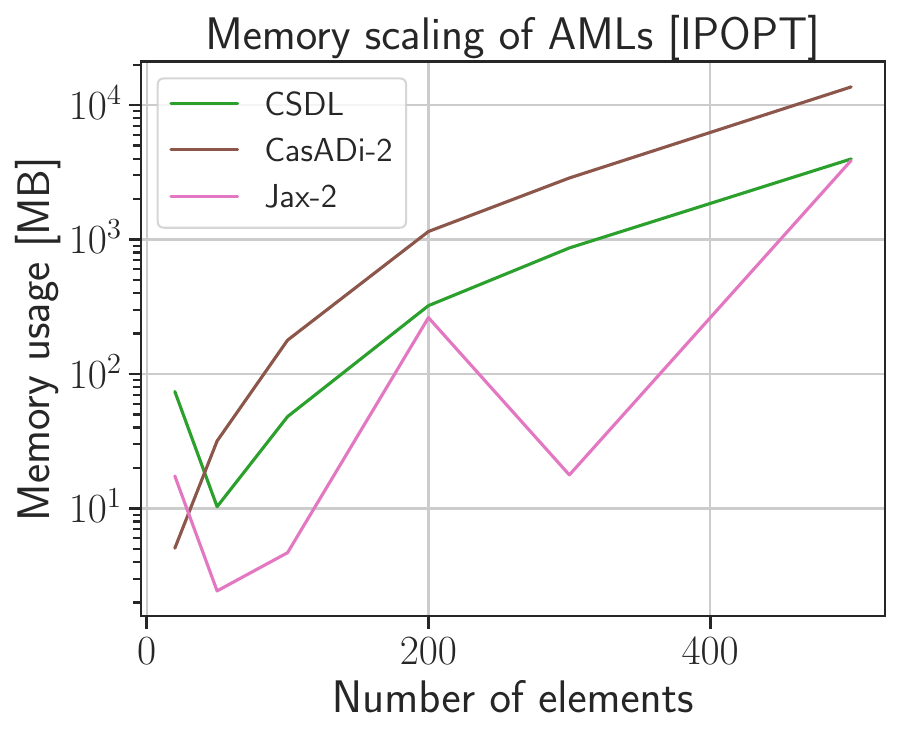}
        % \caption{\textbf{Cantilever beam with point loading}
        %     % \normalfont{}
        % }
        \includegraphics[width=0.45\linewidth]{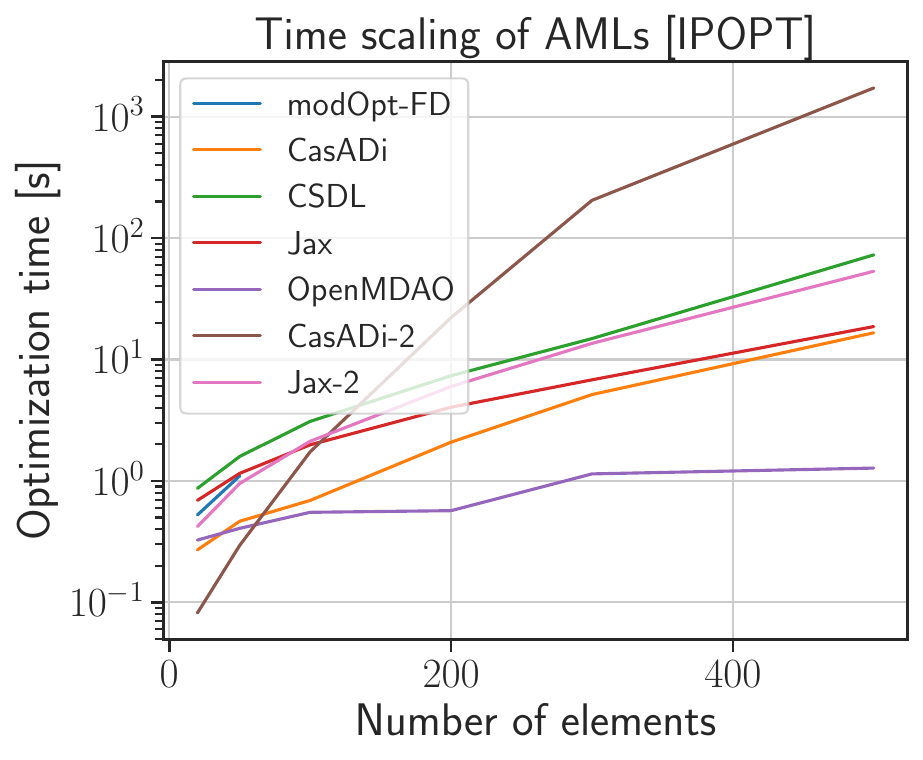}
        \caption{\textbf{Memory usage and solve time comparison of various modeling environments in a 
            scaling study using the cantilever beam problem.}
            The optimizer used is IPOPT.
            % \normalfont{}
        }
        \label{fig:cantilever_amls}
    \end{figure}
    
    We now examine the time and memory scaling of different modeling frameworks
    using IPOPT as the optimizer.
    Memory costs for first-order models across all frameworks, except CSDL, are negligible
    and are therefore omitted from the memory plots in Fig. \ref{fig:cantilever_amls}.
    The results show that the CSDL models and the second-order models implemented in 
    Jax and CasADi scale in memory between quadratic and cubic rates.
    However, CasADi's second-order models are approximately three times more
    memory-intensive compared to those from Jax and CSDL.
    Given that CSDL is a relatively new framework, we anticipate improvements 
    in its memory efficiency in the near future.
    
    In terms of computational time, OpenMDAO demonstrates the fastest performance, 
    surpassing Jax and CasADi by more than an order of magnitude.
    Among the second-order models, CasADi exhibits poor scaling compared to Jax, 
    with performance approximately two orders of magnitude slower for 
    higher-dimensional problems.
    This discrepancy may be attributed to CasADi's inefficiency in obtaining 
    second derivatives through the \textit{for loops} used for computing
    the stiffness matrix, which scales with the number of beam elements.
    
    \subsubsection*{Spacecraft optimal control}
    In this two-dimensional trajectory optimization problem, 
    we aim to optimize the states $\Vec{x}$ and controls $\Vec{u}$ 
    of a spacecraft during its landing phase.
    The state vector is defined as 
    $\Vec{x} = [x, \Dot{x}, y, \Dot{y}, \theta, \Dot{\theta}]$,
    where $x$ and $y$ represent the Cartesian coordinates, 
    $\Dot{x}$ and $\Dot{y}$ denote the corresponding velocities, and
    $\theta$ and $\Dot{\theta}$ are the rotational angle and angular velocity,
    respectively.
    The control vector $\Vec{u} = [T, \beta]$ represents the magnitude and orientation
    of the spacecraft's thrust output.
    The dynamics of the spacecraft are governed by 
    \begin{equation}
        \Dot{\Vec{x}} = f(\Vec{x}, \Vec{u}) =
        \begin{bmatrix}
              \Dot{x} \\
              -\frac{T}{m} \sin{(\beta+\theta)} \\
              \Dot{y} \\
               \frac{T}{m} \cos{(\beta+\theta)} - g \\
              \Dot{\theta} \\
              -\frac{Tl}{2I} \sin{\beta}
        \end{bmatrix},
    \label{eq:starship_dynamics}
    \end{equation}
    where $m$, $l$, and $I$ represent the mass, length, and moment of inertia 
    of the spacecraft, and $g$ denotes the gravitational acceleration.
    The values of these constants are: $m=100,000 \; kg$, $l=50 \; m$, 
    and $g=9.807 \; m/s^2$.
    The moment of inertia is computed as $I=\frac{1}{12}mL^2$, approximating the
    spacecraft as a uniform rod.
    
    We discretize time into $n_t$ steps over a total time of $t=16 s$.
    The states and controls at each timestep form the 
    optimization variables $X\in \Real^{n_t \times 6}$ and $U\in \Real^{n_t \times 2}$.
    The resulting discretized equations lead to the nonlinear programming problem
    \begin{equation}
        \begin{array}{r l}
            \text{minimize}        & \sum_{i=1}^{n_t} U_{i,1}^2 + \sum_{i=1}^{n_t} U_{i,2}^2 
                                   + \sum_{i=1}^{n_t} X_{i,6}^2     \\
            % \min \limits_{h}        & F^T d                 \\
            \text{with respect to} & X, U \\
            \text{subject to}      & X_{1,:} = \Vec{x_0}, \; X_{n_t,:}  = \Vec{x_f}, \\
                                   & 0.4 T_{max} \leq U_{:, 1} \leq T_{max}, \\
                                   & -20^0 \leq U_{:, 2} \leq 20^0, \\
                                   & X_{i+1,:} - X_{i,:} = f(X_{i,:}, U_{i,:}) dt 
                                   \; \; \forall \; \; i \in \{1,2,..., n_t-1\}
        \end{array}
        \label{eq:starship}
    \end{equation}
    where $T_{max}=2210 kN$ is the maximum available thrust,
    $\Vec{x_0} = [0,0,1000,-80,\pi/2,0]$ is the initial condition,
    $\Vec{x_f} = [0,0,0,0,0,0]$ is the final condition, and
    $dt=t/n_t$ is the timestep.
    This formulation minimizes the thrust output, nozzle movement, 
    and angular velocity of the spacecraft, while ensuring 
    the satisfaction of initial and final conditions, 
    control variable limits, and the governing dynamics.
    
    \begin{figure}[ht]
        \centering
        \includegraphics[width=0.45\linewidth]{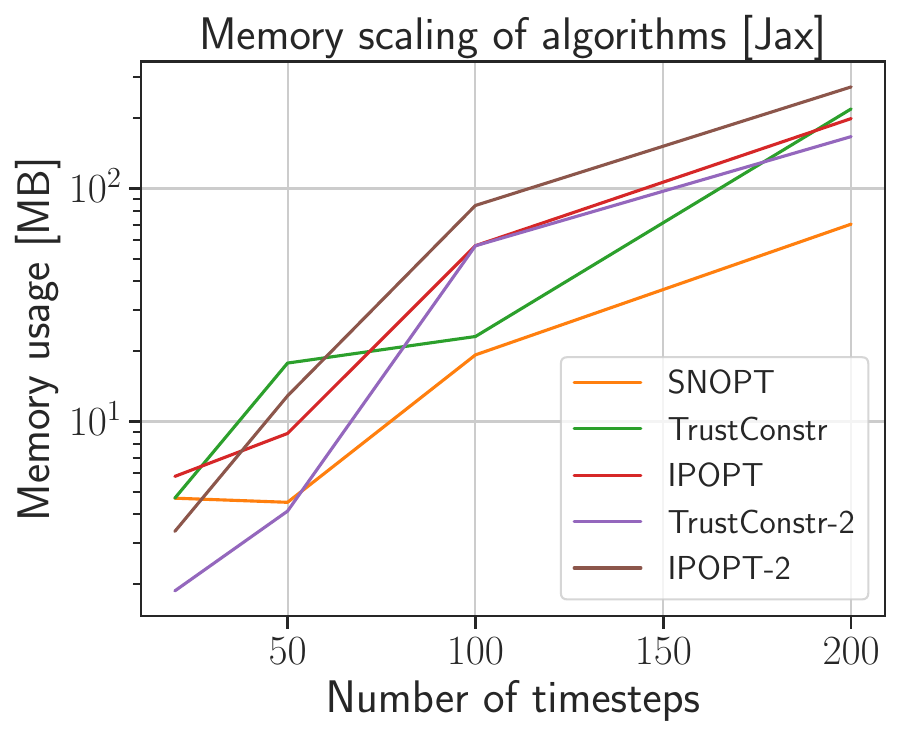}
        \includegraphics[width=0.45\linewidth]{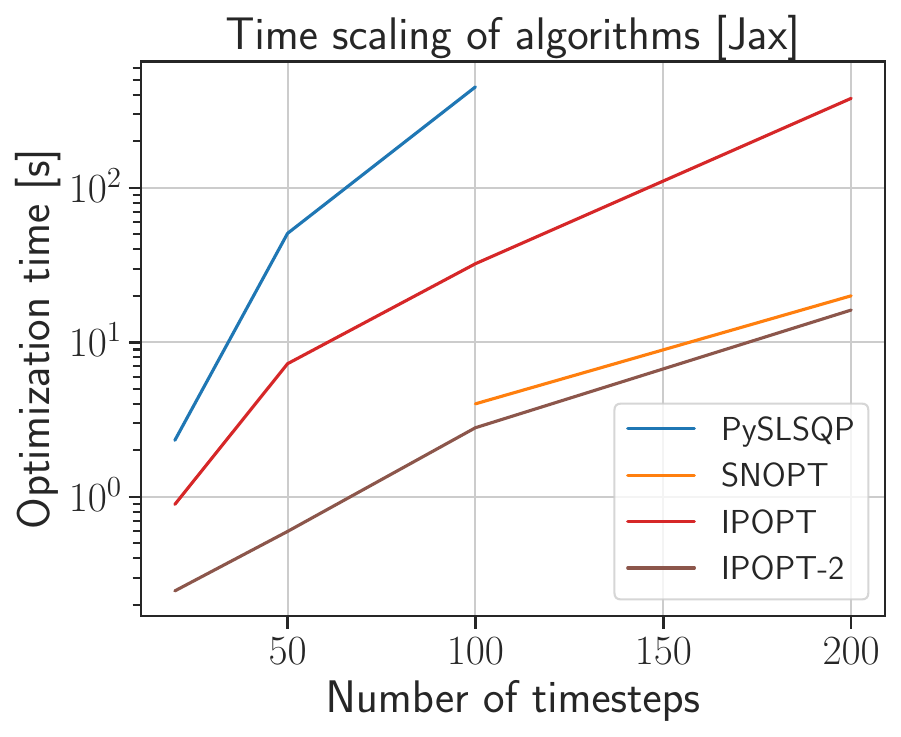}
        \caption{\textbf{Memory usage and solve time comparison of various optimizers in a 
            scaling study using the spacecraft optimal control problem.}
            The models are implemented in Jax.
            % \normalfont{}
        }
        \label{fig:starship_opts}
    \end{figure}
    
    We evaluate the performance of different algorithms on problems 
    discretized with 20, 50, 100, and 200 timesteps.
    Figure \ref{fig:starship_opts} presents a comparison of the memory and
    computational time required by the optimizers.
    In contrast to the cantilever beam problem,
    significant differences emerge between different algorithms, primarily due to
    the larger number of constraints, which scale with the number of timesteps.
    
    PySLSQP consumes negligible memory and is therefore excluded from the memory plot;
    however, its convergence time is an order of magnitude slower than 
    the second slowest method.
    SNOPT fails to converge for coarser discretizations,
    while trust-constr does not converge for any of the problem instances.
    Among the optimizers tested, IPOPT proves to be the most robust, 
    with its second-order variant consistently outperforming the others 
    in terms of convergence time. 
    Notably, when SNOPT successfully converges, 
    its performance closely approaches that of second-order IPOPT,
    while also being the most memory-efficient method.
    
    \begin{figure}[ht]
        \centering
        \includegraphics[width=0.45\linewidth]{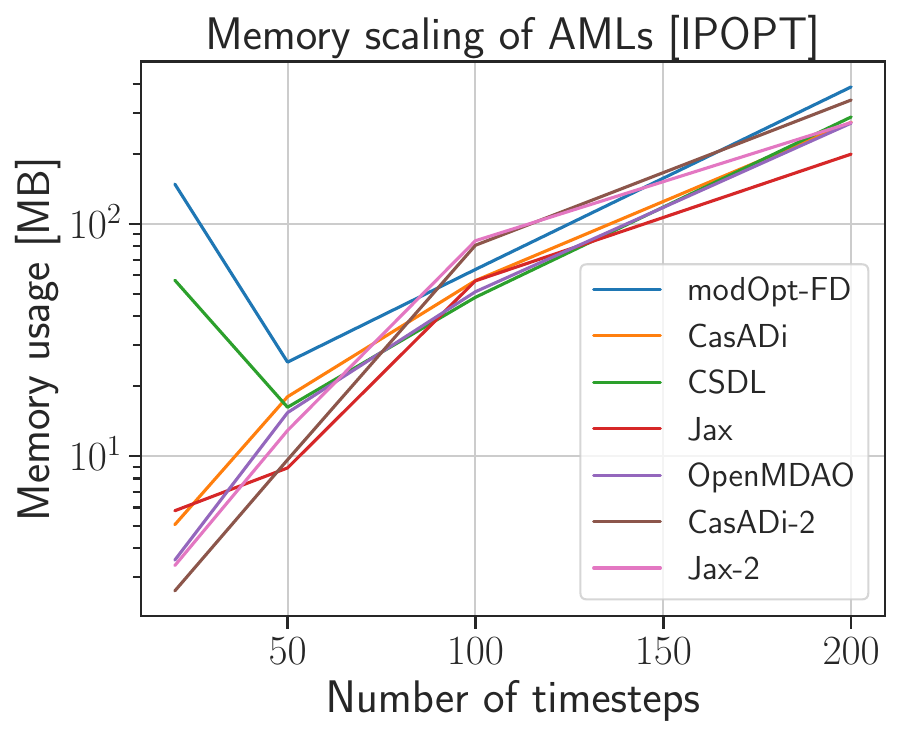}
        \includegraphics[width=0.45\linewidth]{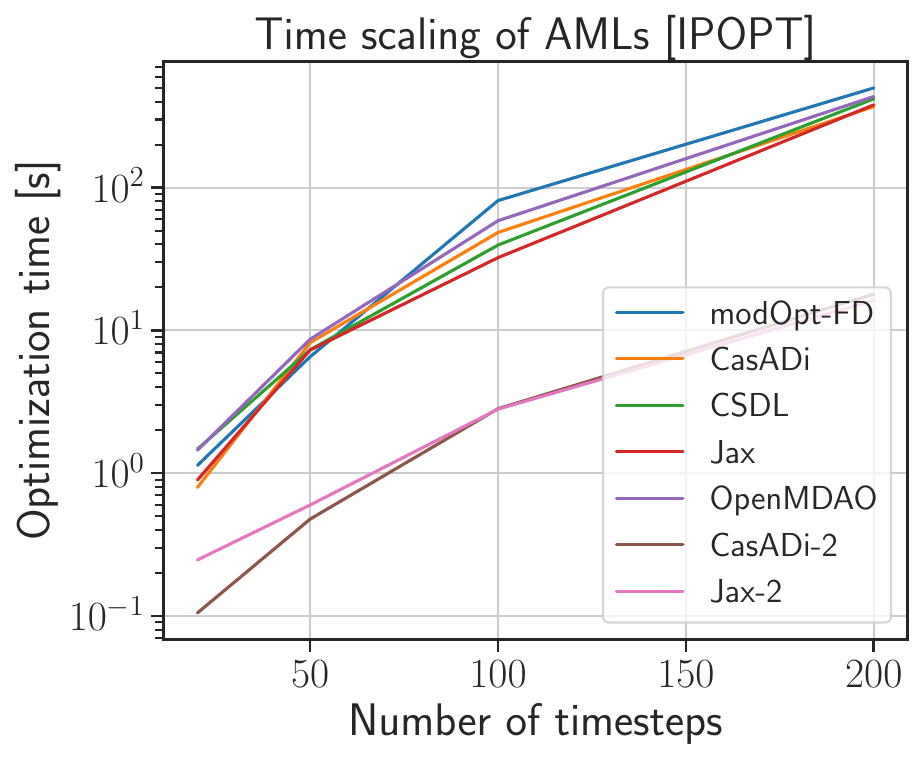}
        \caption{\textbf{Memory usage and solve time comparison of various modeling environments 
            using the spacecraft optimal control problem.}
            The optimizer used is IPOPT.
            % \normalfont{}
        }
        \label{fig:starship_amls}
    \end{figure}
    
    Figure \ref{fig:starship_amls} illustrates the memory and time scaling behavior of 
    modeling frameworks when using IPOPT as the optimizer.
    We do not observe any significant differences in memory usage across models.
    However, second-order models that utilize Hessians result in significantly faster 
    optimization---by more than an order of magnitude---compared to first-order models.
    
    \begin{figure}[ht]
        \centering
        \includegraphics[width=0.45\linewidth]{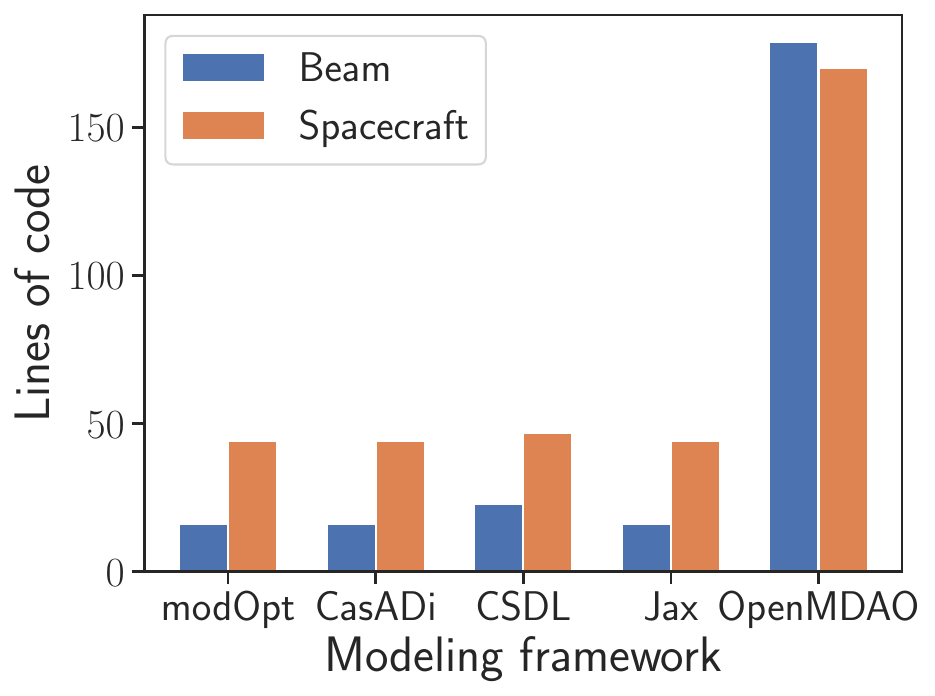}
        \caption{\textbf{Comparison of number of lines of code across different modeling frameworks 
                using the beam and spacecraft models.}
                The count is based on code statements rather than the total number of lines of code 
                to ensure a fair comparison.
            % \normalfont{}
        }
        \label{fig:loc}
    \end{figure}
    
    To this point, we have focused solely on the performance of various modeling
    frameworks without considering the effort required to construct the models.
    Figure \ref{fig:loc} illustrates the number of lines of code (LOC) 
    required to develop the models in different frameworks.
    CasADi, modOpt (using finite difference derivatives), and Jax  
    models require the fewest lines of code, while CSDL necessitates 
    slightly more code than the aforementioned models.
    A key observation is that OpenMDAO requires approximately 5 to 10 times 
    more code statements compared to other frameworks.
    This increased code volume in OpenMDAO arises from the need to decompose models 
    into smaller components and provide partial derivatives for these components.
    Consequently, the additional time and effort required to develop models 
    using OpenMDAO are proportional to the increase in the number of lines of code.
    
    Finally, we conclude this section by summarizing the results from both case studies.
    Among the modeling languages examined, Jax emerges as the most efficient, 
    demonstrating superior performance in terms of both memory usage and 
    computational time, while also requiring the least effort to model 
    moderately complex problems.
    Regarding optimizers, consistent with the No Free Lunch (NFL) theorems, 
    no single optimizer emerges as the definitive best across all scenarios.
    However, several key observations can be made.
    Second-order methods do not necessarily lead to faster optimization for
    large and complex models with implicit state variables; 
    in some cases, they may even slow down the process.
    Conversely, for problems characterized by simple explicit equations involving a large 
    number of variables and constraints, second-order methods have the potential 
    to significantly accelerate optimization, achieving improvements of 
    more than an order of magnitude.
    
    \subsection{Benchmarking with CUTEst}
    \label{sec:cutest_bm}
    
\begin{figure}[ht]
        \centering
        \includegraphics[width=0.45\linewidth]{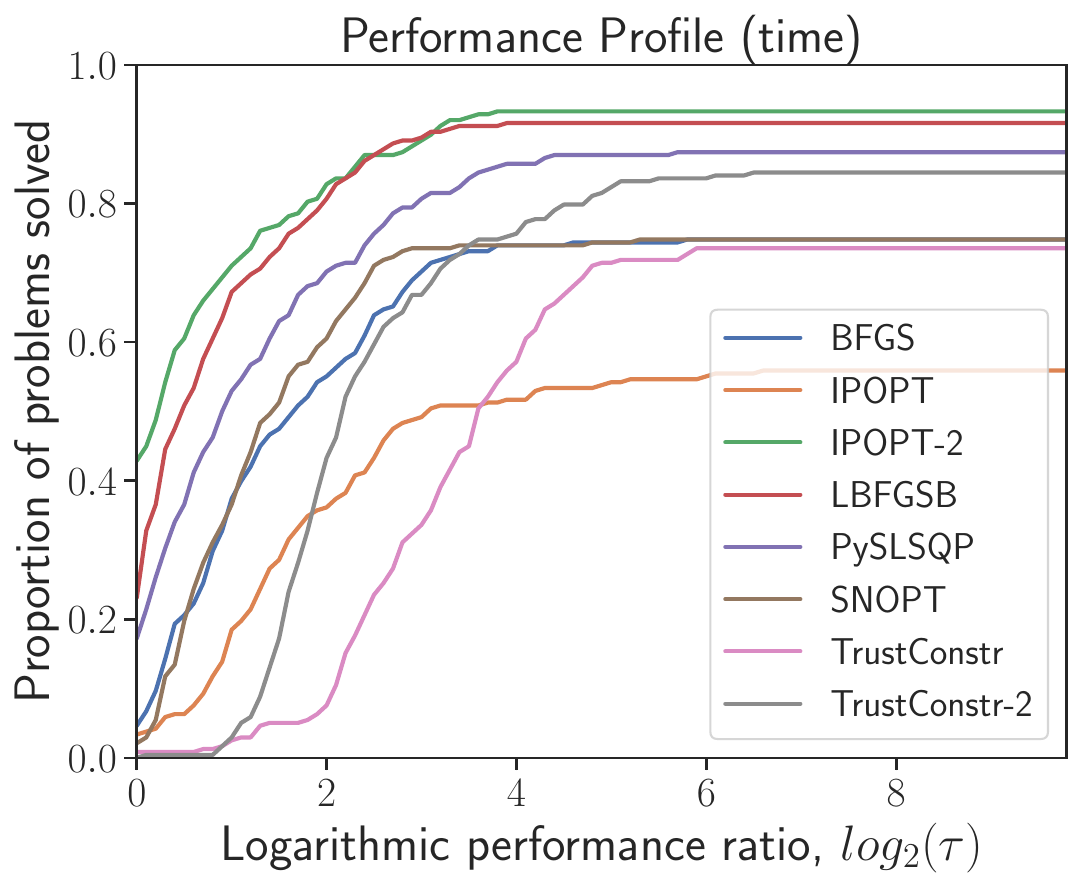}
        \includegraphics[width=0.45\linewidth]{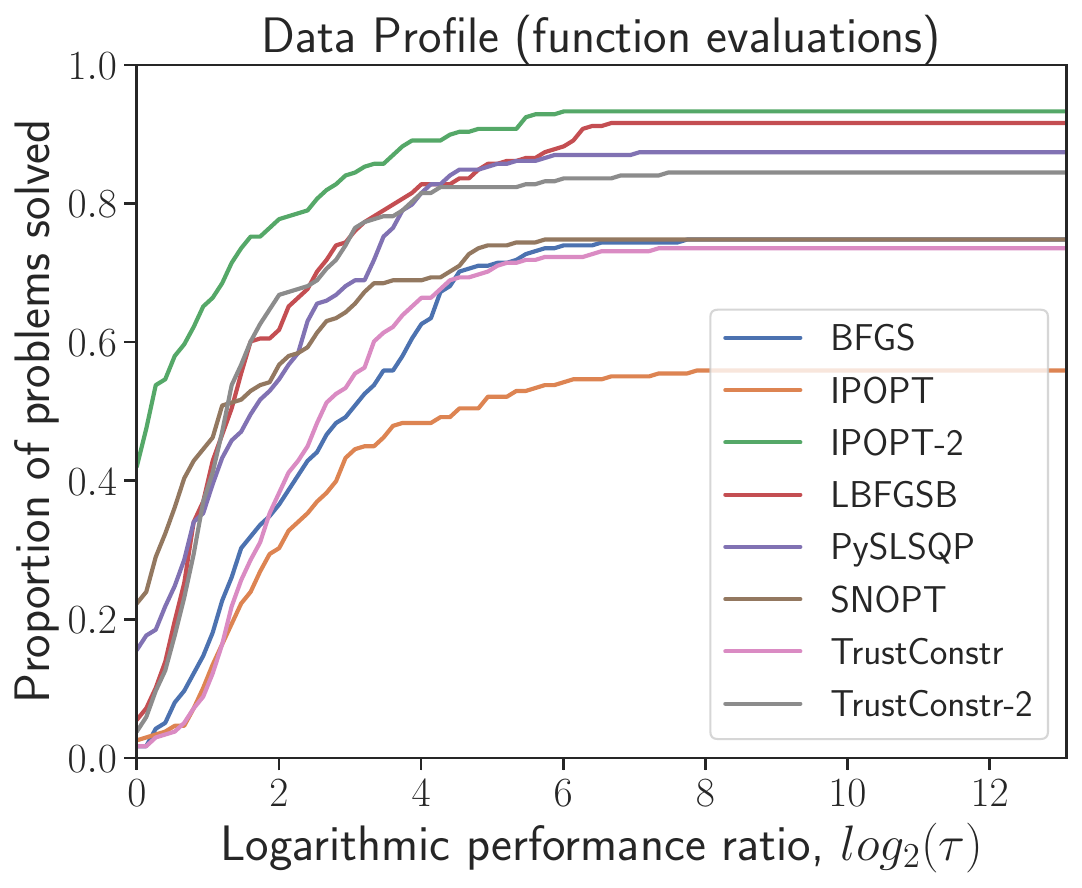}
        \caption{\textbf{Performance and data profiles of various optimizers on 
        unconstrained CUTEst test problems.} The performance profiles are based on time in
        seconds, and the data profiles consider only objective function evaluations.
        The x-axes of the profiles are log-scaled with base 2.
            % \normalfont{}
        }
        \label{fig:performance_cutest}
    \end{figure}

    In this section, we demonstrate the benchmarking enabled by the CUTEst 
    interface in modOpt by solving a set of 238 unconstrained problems with 
    up to 100 variables from the CUTEst test collection.
    Figure \ref{fig:performance_cutest} compares the performance of optimizers using 
    the performance and data profiles generated by modOpt.
    The distinction between these profiles lies in the interpretation of the 
    horizontal axis.
    In a performance profile \cite{more2009benchmarking}, 
    the plot shows the proportion of problems successfully solved by an optimizer
    within a factor $\tau$ of the optimization time required by 
    the best-performing optimizer.
    In contrast, the horizontal axis in a data profile \cite{dolan2002benchmarking} 
    tracks the performance based on the number of function evaluations, 
    rather than the time required by the solvers, 
    making it particularly useful for benchmarking derivative-free methods 
    or problems where model evaluations are computationally expensive.
    
    We initialize all optimizers with their default parameters. 
    A problem is considered successfully solved if the optimizer converges
    within an iteration limit of 500.
    The results obtained largely align with the findings 
    from the coupled Rosenbrock problem discussed earlier.
    
    The profiles indicate that second-order IPOPT is the best-performing optimizer
    for moderate-sized unconstrained problems.
    Despite being first-order methods, both L-BFGS-B and PySLSQP closely 
    follows second-order IPOPT in both performance and data profiles.
    Notably, both second-order IPOPT and trust-constr significantly outperform 
    their first-order counterparts, suggesting that higher-order derivative information 
    is critical for efficient optimization, even for moderate-sized problems.
    
    Among the optimizers tested, first-order IPOPT exhibits the lowest success rate, 
    followed by first-order trust-constr. 
    This suggests that when only first derivatives are available,
    Sequential Quadratic Programming (SQP) methods tend to have a higher likelihood 
    of successful optimization compared to interior point (IP) methods.
    Interestingly, SNOPT and the basic BFGS method using inverse updates
    show comparable performance for the set of unconstrained problems considered.
    
    It is important to note that these trends may shift for larger-scale problems.
    While methods such as PySLSQP and BFGS perform well for small- to medium-sized problems,
    they are known to scale poorly when the number of variables exceeds a few hundred.
    As a result, alternative methods may become more favorable for 
    high-dimensional problems.
    
    % % \input{Body/applications}
    % \input{conclusion}
    \section{Conclusion}
    \label{sec:conclusion}
    
    We introduced modOpt, a modular optimization algorithm development framework designed 
    to support new developers and students in incrementally building or modifying
    optimization algorithms.
    modOpt enables users to assemble optimization algorithms from self-contained modules.
    It is a transparent and flexible platform based on Python and is available 
    open-source on GitHub with documentation.
    
    modOpt's versatility makes it valuable not only for educational purposes 
    but also for advanced developers and practitioners in optimization.
    The framework's built-in interfaces to well-established optimization algorithms 
    such as SNOPT and IPOPT,
    along with interfaces to the most modern modeling languages like CSDL and Jax, 
    and access to the CUTEst test-suite,
    provide users with a robust environment for testing and benchmarking new algorithms.
    
    This paper delved into the software architecture of modOpt,
    discussed the theory behind the algorithms it supports, and
    demonstrated its usage and capabilities.
    We also presented several numerical studies showcasing the unique benefits of modOpt.
    While the primary focus of the paper was on nonlinear programming problems,
    modOpt is designed to integrate algorithms for a broader range of 
    optimization problem classes.
    
    Potential directions for future development of modOpt include 
    support for parallel function evaluations, particularly for 
    finite-difference derivative computation and population-based optimizers,
    the integration of additional modules such as linear system solvers and 
    limited-memory Hessian approximations, and 
    the expansion of the benchmark suite to cover recent optimization applications 
    such as those in machine learning, and large-scale and multidisciplinary
    design optimization.

    \section*{Acknowledgments}
    This material is based upon work supported by the National Science Foundation
    under Grant No. 1917142 and by NASA under award No. 80NSSC23M0217.
    
    \section*{Declarations}
    
    \subsection*{Conflict of interest}
    On behalf of all authors, the corresponding author states that there is no conflict of interest.
    
    \subsection*{Replication of results}
    modOpt is hosted as an open-source project on GitHub at 
    \href{https://github.com/lsdolab/modopt}{https://github.com/lsdolab/modopt}.
    All numerical studies and benchmarks presented in this paper are available 
    in the \texttt{examples/} directory of the GitHub repository.
    Comprehensive documentation, including installation instructions, usage guidelines, 
    and API references, is hosted at 
    \href{https://modopt.readthedocs.io/}{https://modopt.readthedocs.io/}.
    
    % For bibliography using biber engine   
    % \section*{References}
    % \printbibliography[heading=none]

    % For bibliography using OLD bibtex engine
    \bibliographystyle{unsrtnat}
    \bibliography{main}

% For bibliography using biber engine   
% \end{xrefsection}

\end{document}